\documentclass[pra,notitlepage,twocolumn]{revtex4-1}

\usepackage{amsmath}
\usepackage{amssymb}
\usepackage{bm}
\usepackage{graphicx}
\usepackage{times}

\usepackage{xcolor}

\usepackage[utf8]{inputenc}
\usepackage{mathrsfs}
\usepackage{mathdots}
\usepackage{physics}

\usepackage[normalem]{ulem}

\begin{document}


\title{Strong optical coupling combines isolated scatterers into dimer}

\author{Alexey~A.~Dmitriev${}^{1}$}
\email{alexey.dmitriev@metalab.ifmo.ru}
\author{Mikhail~V.~Rybin${}^{1,2}$}

\affiliation{$^1$Department of Physics and Engineering, ITMO University, St Petersburg 197101, Russia}
\affiliation{$^2$Ioffe Institute, St Petersburg 194021, Russia}

\date{\today}

\begin{abstract}
We analyze the transition between different coupling regimes of two dielectric rods, which occurs at a \emph{critical distance} between them. The hallmark of strong coupling regime is the peak splitting effect observed in spectra.  Here we comprehensively evaluate the critical distance as a function of the rod permittivity using a number of different approaches. The scattering spectra of the two rods in dependence on the distance demonstrate the weak to strong coupling transition. We start the analysis by introducing a region of a \emph{tidal energy flux} around a single isolated rod (the region is related to the near field) and demonstrate that its effective radius corresponds to the critical distance obtained from the scattering spectra.  Next, we study the eigenfrequencies of the dimer as functions of distance by `diagonalizing' the coupled multipole matrix. In order to find an analytical formula for the critical distance, we consider the problem under several approximations, which yield similar results.
\end{abstract}

\maketitle

\section{Introduction}
The study of metamaterials and optical antennas composed of resonant meta-atoms~\cite{harris2010emergence, novotny2011antennas} requires a deep insight into formation of collective modes in clusters of nanoparticles. Among others so called oligomers~\cite{kuznetsov2016optically} have gained a lot of attention~\cite{filonov2014field, miroshnichenko2012fano, chandel2019mueller, kuznetsov2016optically, yan2015magnetically, yan2015directional, wang2007plasmonic, greybush2017plasmon, alonso-gonzalez2011real} because of the appearance of bright and dark modes~\cite{cao2011optical, gao2018dark}, enhancement of localized electric and magnetic fields~\cite{bakker2015magnetic, albella2013low, boudarham2014enhancing}, Fano resonances with suppression of the scattering cross-section~\cite{filonov2014field, miroshnichenko2012fano, chandel2019mueller, bachelier2008fano, yan2015directional} and effects of strong coupling~\cite{cao2011optical, gunnarsson2005confined, tsai2012plasmonic}.  The simplest oligomer is a dimer, i.e. a two-particle complex, which has been widely studied for dielectric~\cite{bakker2015magnetic,  albella2013low, cao2011optical, yan2015magnetically, boudarham2014enhancing, boudarham2014enhancing, yan2015directional} and plasmonic~\cite{gao2018dark, bachelier2008fano, gunnarsson2005confined, tsai2012plasmonic, schaffernak2018plasmonic} case.  Plasmonic dimers allow a quasistatic approximation being a simpler mathematical description, however they have inherent Ohmic losses~\cite{khurgin2015how}.  In contrast, dielectric oligomers have low losses~\cite{kuznetsov2016optically}, but a considerable phase retardation owing to a bigger particle size complicates their analysis.

In chemistry atoms are said to form a molecule when robust chemical bonds appear between them. This process is accompanied by a hybridization of atomic orbitals into molecular ones. Similarly, in photonics a non-radiative coupling can hybridize normal modes of isolated meta-atoms into collective ones making an optical oligomer~\cite{wang2007plasmonic, geddes2016reviews, schaffernak2018plasmonic}. Thus, the coupling regime governs whether two particles form a dimer (for a strong coupling) or they should be treated as independent scatterers (for a weak coupling).  In particular, both weak and strong coupling regimes between two silicon rods have been demonstrated experimentally as appearance of the TM$_{01}$ Mie resonance splitting in scattering spectra~\cite{cao2011optical}. The strong coupling between a pair of waveguides makes it possible to achieve tunable optical forces~\cite{fernandes2018center, li2009tunable}.  Besides, an insight into coupling regimes between the neighbor structure elements is crucial for magnetic~\cite{rybin2015phase} and electric~\cite{maslova2018dielectric} dielectric metamaterials. However, to the best of our knowledge, conditions of weak-to-strong coupling transition and the relationship between strong coupling and non-radiative nearfield interaction have yet to be established.

In this paper, we study a system consisting of two infinite dielectric rods.  We analyze the transition between different coupling regimes~\cite{cao2015dielectric}, which occurs when the rods are separated by \emph{a critical distance} $d_c$. This distance is evaluated as a function of the rod permittivity. We utilize the Rabi splitting, also known as the Autler--Townes splitting, observed in spectra, to distinguish the strong coupling regime~\cite{cao2015dielectric, gao2018dark, peng2014what, dietz2007rabi}. Simulations of scattering spectra of two rods are performed for different distances to demonstrate the weak-to-strong coupling transition. To explore this physics, we start by introducing a region of \emph{a tidal energy flux} around the single isolated rod (this region is related to the near field) and demonstrate that its effective radius corresponds to the critical distance obtained from the scattering spectra of the two rods. Next we analyze the eigenfrequencies of the two rods as functions of the distance. We obtain them by `diagonalizing' the matrix of coupled multipoles. In order to derive an analytical formula for the critical distance, we limit the consideration to the conventional coupled oscillator model. Our `Hamiltonian' turns out to be energy-dependent~\cite{formanek2004wave, lombard2009many} and the problem fails to admit a non-numerical treatment, so we fall back on different approximations to derive the criteria for the critical distance.

\section{General theoretical model}
\label{sec:model}

\begin{figure}[t]
  \centering
  \includegraphics[]{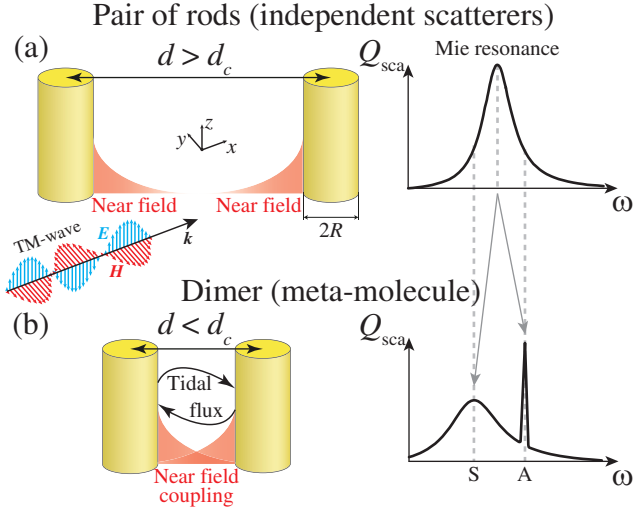}
  \caption{\label{fig:Schematic} 
    Sketch of the system under study.
    (a) Without the nearfield coupling, the two rods scatter light independently.  The scattering spectra demonstrates a single major feature corresponding to the Mie resonance.
    (b) Strong nearfield coupling combines the rods into a dimer. Scattering spectra demonstrate two features, which correspond to symmetric (S) and antisymmetric (A) modes of the dimer. 
  }
\end{figure}

Here we consider a pair of dielectric rods made of a lossless material with permittivity $\varepsilon$, which are infinite along $z$ axis (see Fig.~\ref{fig:Schematic}). We assume no magnetization of the material ($\mu = 1$) so that the refractive index $n = \sqrt{\varepsilon}$. Both rods have the same radius $R$ and the distance between their centers is $d$. Because of the scalability of the Maxwell's equations, the dimensionless size parameter $x = kR$, where $k$ is the wave number, is more convenient than frequency.
We study two polarizations: transverse electric (TE) polarization with the electric field vector being transverse to the rod axis and transverse magnetic (TM) polarization with the electric field being parallel to the rod axis.
Also, throughout the work, we study two geometries of scattering, defined as follows. 
Here we use cylindrical coordinates ($r, \varphi, z$) with the origin placed midway between the rods. Let us choose $\varphi = 0$ along the direction of incident plane wave propagation. Then we define the longitudinal geometry with the rods placed at \(\varphi = 0, r = d/2\) and \(\varphi = \pi, r = d/2\); and the transverse geometry with the rods placed at \(\varphi = \pm\pi/2, r=d/2\).

\subsection{Multiple scattering approach}

We simulate the scattering on the two rods by means of the multiple scattering theory that is a rigorous coupled multipole method, which takes the interaction between all scatterers into consideration~\cite{lloyd1972multiple, leung1993multiple, felbacq1994scattering, nicorovici1995photonic, tayeb2004combined, markos2016photonic, markos2016coupling}.%
The method is based on the point-multipole approximation, which allows us to write the incident ($\psi_i^{(j)}$) and the scattered ($\psi_s^{(j)}$) waves in the vicinity of $j$-th rod as
\begin{equation}
\label{eq:MultipoleDecomposition}
\begin{aligned}
  \psi_i^{(j)} &= \psi_{\mathrm{ext}} + \sum_{l = -N}^N I_{j, l} J_l (k |\mathbf{r} - \mathbf{r}_j|) e^{-il\varphi(\mathbf{r} - \mathbf{r}_j)}, \\
  \psi_s^{(j)} &= \sum_{l = -N}^N S_{j, l} H_l^{(1)} (k |\mathbf{r} - \mathbf{r}_j|) e^{-il\varphi(\mathbf{r} - \mathbf{r}_j)},
\end{aligned}
\end{equation}
  where $\varphi(\mathbf{r})$ denotes the angle of the vector $\mathbf{r}$ in polar coordinates, $k$ is the wavenumber, and $\mathbf{r}_j$ is the rod position. $H_l^{(1)}$ is the Hankel function related to the outgoing waves [here we assume time harmonics of the form $\exp(-i\omega t)$], $J_l$ is the Bessel function related to the incident waves, $N$ is the maximal azimuthal number of multipoles under consideration, and $\psi_{\mathrm{ext}}$ is the excitation field, i.e. the waves form external sources. Scattering on a rod is described by the Lorenz--Mie coefficients: 
\begin{equation}
  \label{eq:LM}
  S_{j, l} = a_{l}I_{j, l},
\end{equation}
where $a_l$ is the Lorenz--Mie coefficient for the $l$-th multipole. As in our problem all rods are equivalent, scattering on each of them is described by the same set of Lorenz--Mie coefficients
\begin{equation}
  a_{l}(\varepsilon, x) = \frac{pJ_l(x)J'_l(x) - J'_l(x)J_l(nx)}{(H_l^{(1)})'(x)J_l(nx) - pH_l^{(1)}J'_l(x)},
\end{equation}
where $p = n$ in case of TM polarization, and $p = -1/n$ in case of TE polarization.

Next we note that the field, which is incident to $j$-th rod, consists of the excitation and the sum of the fields scattered by each other rod:
\begin{equation}
  \label{eq:IncidentField}
  \psi_i^{(j)} = \psi_{\mathrm{ext}} + \sum_{i \neq j} \psi_s^{(i)}.
\end{equation}
In order to rewrite the scattered fields $\psi_s^{(i)}$ as a multipole expansion in the vicinity of the $j$-th rod, we use the formula
\begin{equation}
\label{eq:ReDecompositionFormula}
\begin{aligned}
  H_l^{(1)} (k |\mathbf{r} - \mathbf{r}_a|) e^{-il\varphi(\mathbf{r} - \mathbf{r}_a)} &= \sum_{m = -\infty}^{+\infty} H_m^{(1)} (kr_a) e^{-im\varphi(\mathbf{r}_a)} \times\\&\times J_{l+m}^{(1)} (kr) e^{-i(l+m)\varphi(\mathbf{r})},
\end{aligned}
\end{equation}
where $\mathbf{r}_a$ is an arbitrary vector. 
We apply this formula in order to express the incident multipole amplitudes $I_{j, l}$ through the scattered multipole amplitudes $S_{i, m}$. By substituting the multipole decompositions~\eqref{eq:MultipoleDecomposition} and the formula~\eqref{eq:ReDecompositionFormula} into Eq.~\eqref{eq:IncidentField}, it is straightforward to show that
\begin{equation}
  \label{eq:Decomp}
  I_{j, l} = \sum_{i \neq j} \sum_{m = -N}^{N} H^{(1)}_{l-m}(kr_{ij}) e^{-i(l-m)\phi(\mathbf{r}_{ij})} S_{i, m},
\end{equation}
where $\mathbf{r}_{ij} = \mathbf{r}_i - \mathbf{r}_j$. Here we have truncated the infinite summation at the maximal azimuthal number $N$.

Using the equations~\eqref{eq:LM} and~\eqref{eq:Decomp} together, we can express either $I_{j, l}$ or $S_{j, l}$ (here we choose the latter) to obtain a system of $([2N+1]\cdot M)$ algebraic equations with the same number of unknowns. Here $M$ is the number of rods (in the present study we set $M = 2$).
By solving this system, we can work out the response to any kind of excitation $\psi_{\mathrm{ext}}$, including a plane wave considered here. To do it, we first rewrite the excitation field as a multipole expansion in the vicinity of the $j$-th rod. In the case of a plane wave we use the Jacobi--Anger expansion, which gives
\begin{equation}
  e^{i\mathbf{k}\cdot\mathbf{r}} = \sum_{l = -\infty}^{+\infty} i^l e^{i\mathbf{k}\cdot\mathbf{r}_j} e^{-il\phi(\mathbf{k})} \cdot J_l (k |\mathbf{r} - \mathbf{r}_j|) e^{-il\varphi(\mathbf{r} - \mathbf{r}_j)}.
\end{equation}

The system of the algebraic equations obtained from combining Eq.~\eqref{eq:LM} and Eq.~\eqref{eq:Decomp} can be written as a single matrix equation. The coupled multipole matrix, which consists of $(M\times M)$ blocks of size $([2N+1] \times [2N+1])$, acts on the vector containing the scattered multipole amplitudes $S_{j, l}$. The result is a vector containing the multipole amplitudes of the excitation field. It is straightforward to show that the matrix equation for the case of two rods can be expressed as
\begin{equation}
\label{eq:MSTMatrix}
\left[
\begin{array}{c|c}
E & V(\mathbf{d}) \\
\hline
V(-\mathbf{d}) & E
\end{array}
\right]
\left[
\begin{array}{c}
\vdots \\
S_{1, -1} \\
S_{1, 0} \\
S_{1, +1} \\
\vdots \\
\hline
\vdots \\
S_{2, -1} \\
S_{2, 0} \\
S_{2, +1} \\
\vdots \\
\end{array}
\right] =
\left[
\begin{array}{c}
\vdots \\
i^{-1}e^{i\mathbf{k}\cdot\mathbf{r}_1} e^{i\phi(\mathbf{k})} \\
e^{i\mathbf{k}\cdot\mathbf{r}_1} \\
ie^{i\mathbf{k}\cdot\mathbf{r}_1} e^{-i\phi(\mathbf{k})} \\
\vdots \\
\hline
\vdots \\
i^{-1}e^{i\mathbf{k}\cdot\mathbf{r}_2} e^{i\phi(\mathbf{k})} \\
e^{i\mathbf{k}\cdot\mathbf{r}_2} \\
ie^{i\mathbf{k}\cdot\mathbf{r}_2} e^{-i\phi(\mathbf{k})} \\
\vdots \\
\end{array}
\right]
\end{equation}
where $E$ is the $([2N+1]\times[2N+1])$-sized unit matrix, and
$V_{l,m}(\mathbf{d}) = -a_lH^{(1)}_{l-m}(kd)e^{-i(l-m)\phi(\mathbf{d})}$.
Zeros of the determinant of the coupled multipole matrix are essentially the eigenfrequencies of the two rods.  The inverse coupled multipole matrix being an analog of a Green function allows us to solve the electrodynamic problem.

The field $\psi_s$ scattered by the two rods is expressed as follows
\begin{equation}
  \label{eq:ScaField}
  \psi_s = \sum_{j = 1}^M\sum_{l = -N}^N S_{j, l}
  H_l^{(1)} (k |\mathbf{r} - \mathbf{r}_j|)
  e^{-il\varphi(\mathbf{r} - \mathbf{r}_j)}.
\end{equation}

By considering the far-field asymptotic of the scattered field~\eqref{eq:ScaField} along the direction of incidence, we express the forward scattering amplitude
\begin{equation}
\label{eq:Es}
f(0) =
\sqrt{\frac{2}{\pi k}}
\sum_{j=1}^M\sum_{l = -N}^{N}
  \left[
    i^{-l}
    e^{-i\frac{\pi}{4}}
    e^{-i\mathbf{k}\cdot\mathbf{r}_j}
  \right]
  \frac{S_{j,l}}{\abs{\psi_{\mathrm{ext}}}}.
\end{equation}
This expression is utilized to evaluate the extinction cross-section by means of the 2D optical theorem~\cite{dmitriev2018coupling, mitri2015optical, gu1989some}
\begin{equation}
  \sigma_{\mathrm{ext}} = -2\sqrt{\frac{\pi}{k}}\left(\Re{f(0)} - \Im{f(0)}\right).
\end{equation}

As follows from the symmetry, the transverse scattering geometry forbids the excitation of antisymmetric modes, while the longitudinal geometry does not.  Thus, we can distinguish between symmetric and antisymmetric modes by comparing the scattering spectra in the longitudinal and transverse geometries.

\subsection{Simulation results}
\label{sec:simulation}

\begin{figure}[t]
  \centering
  \includegraphics[]{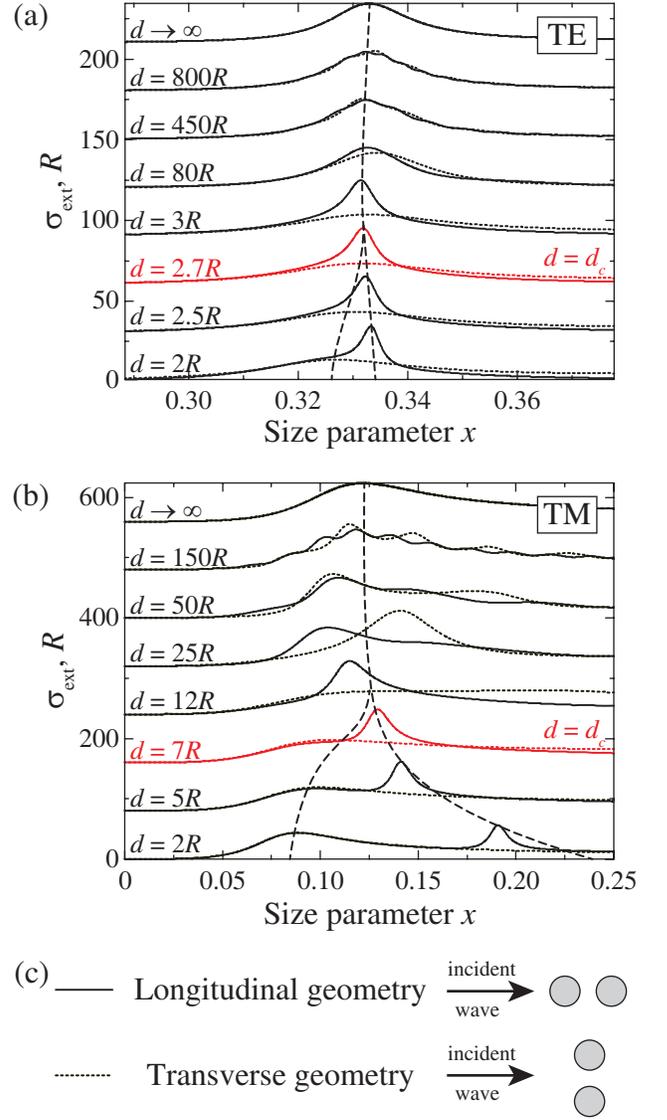}
  \caption{\label{fig:spectra} Scattering spectra of two infinite dielectric rods as a function of distance for TE (a) and TM (b) polarizations. The rods are made of a material with $\varepsilon = 50$. Panel (c) shows the legend for both plots (a) and (b). The solid lines show the extinction cross-section for the longitudinal scattering geometry that allows the excitation of antisymmetric modes. The dotted lines show the extinction cross-section for the transverse geometry that forbids it. The dashed lines are guides for the eyes only. The spectra are relatively shifted along the vertical axis.
  }
\end{figure}

In Fig.~\ref{fig:spectra} we plot the scattering spectra $\sigma_{\mathrm{ext}}$ for both TE and TM polarizations simulated for two rods with $\varepsilon = 50$ (e.\,g., distilled water in the microwave range \cite{andryieuski2015water}) by taking 7 multipoles ($\abs{m} \le 3$) into account. 
We identify three coupling regimes dependent on the distance between rods. 
At almost infinite distance between the rods (here we use $L = 10^6 R$) the scattering spectra of the two rods mimic that of a single isolated rod, with the only peak corresponding to the dipole Mie resonance (the curves are labeled as $L \to \infty$ in Fig.~\ref{fig:spectra}). This means the rods scatter the incident plane wave independently, i.e. the coupling between the rods is infinitesimal.

The spectra labeled $d = 450R$ and $d = 800R$ for TE polarization, as well as $d = 25R$, $d = 50R$ and $d = 150R$ for TM polarization, exhibit weak fringes, i.e. oscillations.  They are approximately equidistant, with the `period' decreasing as the distance $d$ increases.  These fringes correspond to Fabry--Perot-like eigenmodes present in the considered system due to the coupling via quasi-free waves traveling between the rods.  We also note that in the transverse geometry the `period' of fringes is twice as large as in the longitudinal geometry, since the transverse geometry forbids excitation of the antisymmetric modes and only the peaks that correspond to symmetric modes remain.
We estimate the lowest Fabry--Perot frequency that corresponds to the half-wavelength equal to the distance between rods. When it is much higher than the frequency $x_0$ of the lowest Mie resonance ($\pi R/d \gg x_0$), the fringes are not observed in the examined spectral range.  This estimation is in a good agreement with plot in Figure~\ref{fig:spectra} where the fringes corresponding to Fabry--Perot-like modes are absent for the distances $d \leqslant 12R$ in TM~polarization and for $d \leqslant 80R$ in TE~polarization.

At distances less than $d_c \simeq 10R$ for TM~polarization and less than $d_c \simeq 2.75R$ for TE~polarization, the spectra in the longitudinal geometry demonstrate a splitting of the dipole peak corresponding to the symmetric mode with a low quality ($Q$) factor and the high-$Q$ antisymmetric mode.  In the transverse geometry, which forbids the excitation of the antisymmetric mode, only the low-$Q$ peak corresponding to the symmetric mode is present.  Such kind of a resonance splitting with the formation of two common modes is the hallmark of the strong coupling regime~\cite{cao2015dielectric, gao2018dark}.

\begin{figure}[t]
  \centering
  \includegraphics[]{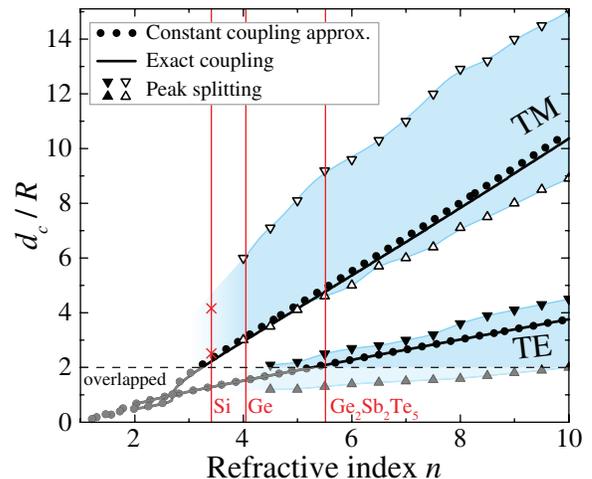}
  \caption{\label{fig:CriteriaComparison}
  Critical distances between rods in TE and TM polarizations corresponding to weak-to-strong coupling regime transition according to various criteria.
  The triangles show data obtained from simulated scattering spectra;
  the solid lines correspond to Eq.~\eqref{eq:NumericalCriterion} (the first crossing of the dimer eigenfrequencies sought for numerically with exact expression for coupling);
  the dotted lines correspond to Eq.~\eqref{eq:ConvCoupling} (the criterion obtained by generalization of the conventional coupled oscillator model and approximating the coupling as constant). The dashed horizontal line divides physical boundary of the multiple scattering theory applicability (non-overlapped rods at $d>2R$).
Red vertical lines show refractive index values of several optical materials in near infrared (Si, Ge, Ge$_2$Sb$_2$Te$_5$ \cite{baranov2017all,wuttig2017phase}). The red crosses values correspond to the experiments with silicon nanowires \cite{cao2011optical}.
  }
\end{figure}

The dependence of the critical distance on the refractive index $d_c(n)$ can be evaluated by analyzing the scattering spectra. Unfortunately, this analysis does not provide the exact value of the distance where the peak splitting appears. However, we can distinguish two extreme cases. (i)~The splitting has definitely appeared when the spectrum  demonstrates two peaks and there is a dip between them (see, Fig.~\ref{fig:spectra}~(b), $d = 5R$). (ii)~The spectrum demonstrates a single intensive peak that has a symmetric lineshape (see Fig.~\ref{fig:spectra}~(b), $d = 12R$), i.e., before a weak hump corresponding to the second peak appears on either side (see Fig.~\ref{fig:spectra}~(b), $d = 7R$). We use these two cases as an error bar for the critical distance obtained from the scattering spectra. Fig.~\ref{fig:CriteriaComparison} shows the distances corresponding to a single `hump-less' peak with down-facing triangles and the distances corresponding to appearance of a dip between the two peaks with up-facing triangles. For both polarizations the critical distance increases linearly with the refractive index of the rods. 

We have also marked in Fig.~\ref{fig:CriteriaComparison} the region of the weak-to-strong coupling transition that has been observed experimentally in silicon nanowires by Cao et al.~\cite{cao2011optical}. The region of the transition was extracted from the scattering spectra in the same manner as we did for the simulated spectra. We note the good agreement of the experimental data with the results of our simulations.

\section{Energy transfer analysis}
\label{sec:EnergyTransfer}

In this section we demonstrate the link between the peak splitting effect and the appearance of a non-radiative energy exchange.
First, we prove that the farfield coupling cannot lead to the splitting of the dipole resonance into a duplet of the symmetric and the antisymmetric modes.  The sketch of the spectra in~Fig.~\ref{fig:Schematic} shows that Mie scattering is weak at the frequencies of the symmetric and the antisymmetric modes due to the frequency mismatch. Thus, at these frequencies the backscattering is weak, and a resonance created by multiple scattering on the rods (as in the Fabry--Perot-like modes) cannot be intensive. Therefore, the farfield coupling cannot lead to the appearance of two intensive peaks of scattering at the frequencies of the symmetric and the antisymmetric eigenmodes, which means this effect is attributed to the near field.

\begin{figure*}[tb]
\includegraphics[]{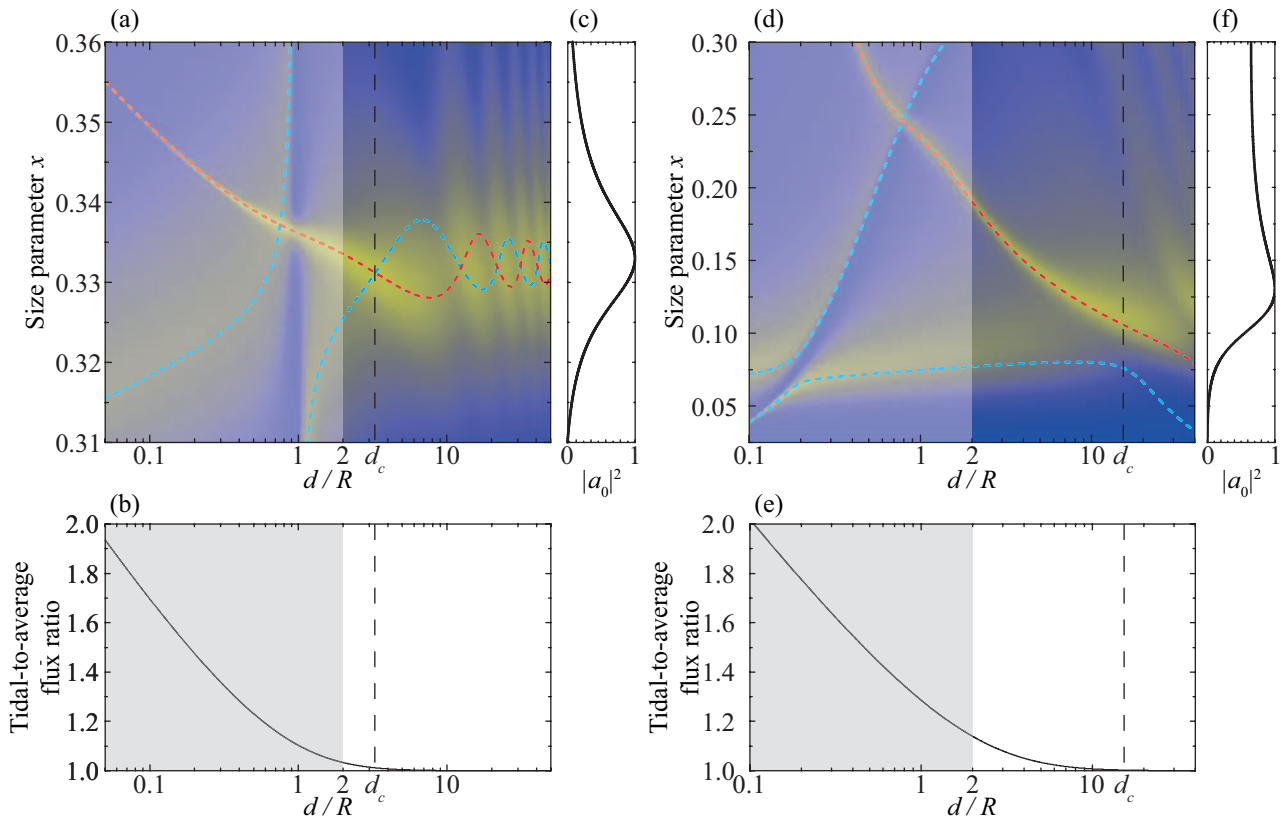}
\caption{\label{fig:colormap}
  (a, d) Scattering spectra of two rods with $\varepsilon = 50$ as functions of distance in TE (a) and TM (d) polarizations, obtained by the multiple scattering approach with dipole and quadrupole taken into account. The dashed lines show the dimer eigenfrequencies obtained numerically by performing a `diagonalization' of the coupled multipole matrix. The non-physical distances $d < 2R$ are shaded in gray.
  (b, e) The tidal to time-averaged energy fluxes ratio around a single isolated rod with $\varepsilon = 50$, as functions of distance, in TE (b) and TM (e) polarizations.
  (c, f) Lineshape of the dipole Mie resonance of a single isolated rod with $\varepsilon = 50$ in TE (c) and TM (f) polarizations.
}
\end{figure*}

In the 3D case nearfield and farfield components of a field generated by a dipole source enter the expression as different terms and can easily be separated. In contrast, the 2D case, which is studied here, provides no means to separate them explicitly.
Thus, in order to distinguish between the far and near field effects rigorously, we analyze instantaneous and time-averaged energy flux around a rod.
The time-averaged energy flux is carried entirely by the farfield component of the wave regardless of the distance from the source (see Appendix~\ref{sec:AppendixA} for a rigorous proof in the 2D case).
Therefore, the nearfield component does not participate in averaged flux, however the instantaneous energy flux associated with it might be great. To evaluate the energy carried by the near field, we introduce \emph{a tidal energy flux} as the time-averaged absolute value of the instantaneous energy flux $\langle\abs{S(t)}\rangle$. We consider the tidal-to-average ratio of the energy fluxes as a figure-of-merit of the energy carried by the near field.

To analyze the energy flux, we adopt the dipole approximation.
The electric and magnetic fields outside a rod are expressed by the Hankel function and its derivative (see Appendix~\ref{sec:AppendixA}). We evaluate the instantaneous Poynting vector using the asymptotic expansions of the Hankel functions near zero (see, e.g., Ref.~\onlinecite{watson1995treatise}), truncated after the first term.  By neglecting the constant terms in favor of $\ln(kr)$, we get
\begin{equation}
  \langle\abs{S_r}\rangle = -\abs{\langle S_r \rangle} \qty(\frac{2}{\pi})^2
  \ln(kr).
\end{equation}
Thus, the tidal-to-average ratio approaches unity in the farfield region and decreases almost linearly with $\ln(kr)$ in the nearfield region.

We plot the dependence of the tidal-to-average ratio on the logarithm of distance in Fig.~\ref{fig:colormap}.
Its linear decrease indicates the nearfield region, while approaching unity indicates the farfield region. 
It is instructive to compare it with the scattering spectra shown in Fig.~\ref{fig:colormap}a,d. The transition from constant to linear behavior of the flux ratio coincides with the splitting of the dipole peak observed in the spectra. It confirms the strong coupling to be an inherently nearfield effect.

\section{Eigenmode analysis}

\begin{figure*}[t]
  \includegraphics[]{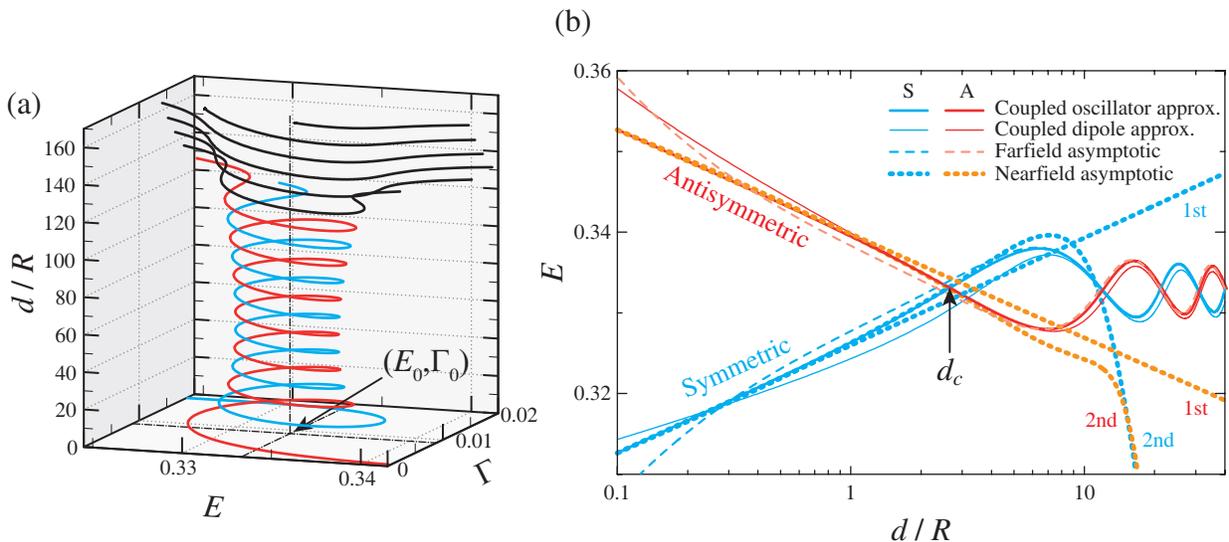}
  \caption{\label{fig:Eigenmodes}
  Eigenfrequencies of two rods with $\varepsilon = 50$ in TE~polarization.
  (a) Complex eigenfrequencies obtained in the dipole approximation, including the Fabry--Perot-like eigenmodes.
  (b) Real parts of the dimer eigenfrequencies, computed with the various approximations:
  coupled oscillator approximation (thick solid lines),
  dipole approximation (thin solid lines),
  farfield asymptotic (dashed lines)
  and nearfield asymptotic (dotted lines), truncated at the first (1st) and the second term (2nd).
  }
\end{figure*}
\label{sec:EigenmodeAnalysis}

Although the strong coupling regime can be qualitatively explained by the appearance of non-radiative energy exchange between the subsystems, the quantitative description is usually provided by solving an eigenvalue problem for mode spatial distributions (or wave functions) in a system. In this section we attack the problem by exploiting the rigorous approach and a number of approximations as well.

We note that for optical materials the transition from weak to strong coupling regime occurs near $d = 2R$, especially in the case of TE polarization. For this reason, the picture in the strong coupling regime remains unclear.  Thus, we also consider here the non-physical distances $d < 2R$ that would correspond to overlapping rods for an illustrative purpose and a deeper physical insight into the strong coupling regime.
We notice that the multiple scattering theory exploits the point multipole approximation that does not contain any information about the radii of the rods, so the results for overlapped rods do not contain any peculiarities owing to the change in the topology of the system (only one boundary with air, instead of two such boundaries in case of non-overlapping rods). Therefore, the results that are physically correct correspond to non-overlapping rods $d \ge 2R$. Taking this in mind we also consider continuation of the solutions for non-physical distances ($d < 2R$), as well.

\subsection{Rigorous results}

We have computed the eigenfrequencies of the dimer as zeros of the determinant for the $([2N+1]M)\times ([2N+1]M)$  matrix of coupled multipoles with $N = 2$ (dipole and quadrupole are taken into account). 
Aside of them, there are two distinct eigenfrequencies which oscillate at large distances and demonstrate the appearance of a splitting, referred to as the Rabi splitting in similar quantum mechanical problems, at small distances.
These are the \emph{dimer eigenfrequencies} of our special interest.
We plot them with dashed lines in Fig.~\ref{fig:colormap}a,d. At small distances $d$, the Rabi splitting  increases approximately linearly, as the logarithm of distance decreases.  We notice that in the TE~polarization (see Fig.~\ref{fig:colormap}a) the splitting is symmetric with respect to the Mie resonance frequency, unlike the TM~polarization (see Fig.~\ref{fig:colormap}d) where the frequency corresponding to the antisymmetric mode demonstrates a shift much larger than the symmetric mode. This asymmetry is related to the asymmetry of the Mie resonance lineshape (see Fig.~\ref{fig:colormap}c,f). The dimer eigenfrequencies diverge in a symmetrical manner when the lineshape of the Mie resonance is close to the Lorentzian curve, i.e., is symmetric.
The frequency corresponding to the symmetric dimer mode also demonstrates an avoided crossing, which is due to the dipole--quadrupole interaction.

\subsection{Dipole approximation}

As we are going to study the dipole resonance, we have limited ourselves to the dipole approximation. The coupled multipole matrix~\eqref{eq:MSTMatrix} acting on a vector composed of the dipole amplitudes of the rods reads
\begin{equation}
  \label{eq:CoupledDipoles}
  \left[
  \begin{matrix}
    1 & -a_0(\varepsilon, \xi)H^{(1)}_0(d \xi/R) \\
    -a_0(\varepsilon, \xi)H^{(1)}_0(d \xi/R) & 1 \\
  \end{matrix}
  \right],
\end{equation}
where \(a_0\) is the Lorenz--Mie coefficient and $\xi$ is the complex size parameter (\(\xi = E + i\Gamma = (k' + ik'')R\)).

Fig.~\ref{fig:Eigenmodes}a shows the eigenfrequencies $\xi$ computed numerically for different distances $d$ by finding the zeros of the determinant of the matrix~\eqref{eq:CoupledDipoles} for rods with $\varepsilon = 50$ (TE~polarization). Thin solid lines in Fig.~\ref{fig:Eigenmodes}b show their real parts $E$.  Let us follow the eigenfrequencies starting from the non-physical distances as small as $d = 0.1R$. At small distances the splitting between the real parts of the eigenfrequencies decreases monotonously with the logarithm of the distance. 
At $d = 3.2R$ real parts of the eigenfrequencies coincide. Fig.~\ref{fig:Eigenmodes}a demonstrates that starting from $d = 3.2R$, the complex eigenfrequencies revolve in the complex plane around the eigenfrequency of single rod Mie resonance $(E_0, -\Gamma_0)$. At $d \simeq 130R$ they merge with the eigenfrequencies corresponding to the Fabry--Perot-like modes.

\subsection{Coupled oscillator model}

In order to work out a criterion for the transition from weak to strong coupling regime analytically, we use the conventional coupled oscillator model. To obtain the effective Hamiltonian, we use the first-order Taylor expansion of the reciprocal Lorenz--Mie coefficient $a_0^{-1}(\varepsilon, \xi)$ around the frequency \(\xi \simeq E_0 - i\Gamma_0\) of the dipole Mie resonance. The expansion reads $a_0^{-1} \simeq i(\xi - E_0 + i\Gamma_0)/\Gamma_0$. After substituting it into the coupled dipole matrix~\eqref{eq:CoupledDipoles} we notice that the eigenfrequencies $\xi$ are the eigenvalues of the effective Hamiltonian
\begin{equation}
  \label{eq:CoupledOscillators}
  \mathscr{H} = 
  \left[
  \begin{matrix}
    E_0 - i\Gamma_0 &
    -g \\
    -g &
    E_0 - i\Gamma_0 \\
  \end{matrix}
  \right],
\end{equation}
where $g = -i\Gamma_0H^{(1)}_0(d \xi/R)$ is the coupling constant.  Below we consider the TE~polarization as an example.

The coupling constant $g$ depends on the eigenfrequency $\xi$. Thus, the Hamiltonian of the problem is energy-dependent. We notice that energy-dependent Hamiltonians occur frequently in two-body problems~\cite{lombard2009many}. The equation
\begin{equation}
  \label{eq:CoupledOscillatorsEigenfreqs}
  \qty(\xi - E_0 + i\Gamma_0)^2 + \qty(\Gamma_0
  H^{(1)}_0\qty(d\xi / R))^2 = 0.
\end{equation}
for the eigenvalues $\xi$ of the Hamiltonian is transcendental and has to be solved either numerically, or within some reasonable approximations. This equation has an infinite number of solutions due to the presence of an oscillating Hankel function with an argument containing the unknown $\xi$.

We also note that the Hamiltonian has an infinite number of eigenvalues and only two eigenvectors --- $[1, 1]^{\top}$, which corresponds to all symmetric modes, and $[1, -1]^{\top}$, which corresponds to all antisymmetric modes.  Orthogonality of these modes forbids their interaction.  It explains the absence of the Fano resonance, which is typically observed in the scattering spectra of dimers of spheres and is attributed to the interaction of the electric dipole and magnetic dipole modes~\cite{yan2015magnetically, yan2015directional, kuznetsov2016optically}.

\subsubsection{Numerical solution}

Thick solid lines in Fig.~\ref{fig:Eigenmodes}b show the dimer eigenfrequencies obtained by solving the equation~\eqref{eq:CoupledOscillatorsEigenfreqs} numerically for $\varepsilon = 50$ in the case of TE~polarization.  The eigenfrequencies computed in the coupled oscillator model agree well with those computed in the coupled dipole approximation, which validates the model.  At the distances greater than $d = 3.2R$ the complex eigenfrequencies revolve around the Mie resonance, just like in the coupled dipole model. At the distances less than $d = 3.2R$ the splitting between the real parts of the eigenfrequencies decreases approximately linearly with the logarithm of the distance. At $d = 3.2R$ the imaginary part of the symmetric mode is close to $-2\Gamma_0$ and that of the antisymmetric mode is close to zero.

As is seen from Fig.~\ref{fig:Eigenmodes}b, the eigenfrequencies demonstrate a large splitting before the first crossing. Thus, we can define the critical distance as the distance which corresponds to the first crossing of the real parts of the complex eigenfrequencies. To find the corresponding distance, we solve the equation~\eqref{eq:CoupledOscillatorsEigenfreqs} in the following manner. We notice that at the crossing point the real parts of both dimer eigenfrequencies $\xi_{1, 2}$ are equal to $E_0$:
\begin{equation}
  \xi_{1,2}(d_c) = E_0 + i\Gamma_{1,2}(d_c).
\end{equation}
By substituting this into Eq.~\eqref{eq:CoupledOscillatorsEigenfreqs}, we get a complex equation with two real unknowns $d_c$ and $\Gamma(d_c)$. It can be transformed into a complex equation of the form $f(z) = 0$ with a single unknown if we introduce $z = d_c + i\Gamma(d_c)$. The result reads
\begin{equation}
\label{eq:NumericalCriterion}
\begin{aligned}
    d_c &= \Re{z}, \\
    \qty(\frac{\Im{z}}{\Gamma_0} - 1)^2 &= \qty[
    H^{(1)}_0\qty(\frac{\Re{z}}{R}(E_0 - i\Im{z}))]^2
\end{aligned}
\end{equation}
We solve the equation~\eqref{eq:NumericalCriterion} numerically for different refractive indices in TE and TM polarizations. Fig.~\ref{fig:CriteriaComparison} demonstrates this result to compare it with the data obtained from peak analysis in the scattering spectra. We find it to be in good agreement with the splitting points in both polarizations. 

\subsubsection{Nearfield asymptotic}

The coupling coefficient for the two rods is proportional to the Green function of the 2D Helmholtz equation, i.\,e., a wave induced by a dipole oscillator. Therefore, we can study it from the point of view of the transition from the nearfield coupling to the farfield coupling. To do that, we utilize the asymptotic expansions of the $H_0^{(1)}(\xi)$ function at $\xi \to 0$ and at $\xi \to \infty$.

First, we consider the asymptotic expansion at $\xi \to 0$ (see, e.\,g., Ref.~\onlinecite{watson1995treatise}).  In order to study the nearfield coupling analytically, we truncate this expansion at the first term. By substituting the resulting expression into Eq.~\eqref{eq:CoupledOscillatorsEigenfreqs}, we obtain an analytical equation for the complex eigenfrequencies in the nearfield region:
\begin{equation}
  \label{eq:NearfieldEigfreq}
  (\xi - E_0 + i\Gamma_0)^2 =
  -\Gamma_0^2
  \qty[
    1 + i\frac{2}{\pi}\qty(\ln{\frac{d \xi}{2R} + \gamma})
  ]^2,
\end{equation}
where $\gamma \simeq 0.57$ is the Euler--Mascheroni constant.
This equation has two complex roots which can be evaluated analytically:
\begin{subequations}
\begin{align}
  \label{eq:NearfieldSym}
  \xi_1(d) &= 
  -\frac{2}{\pi} \Gamma_0 W_{-1}\qty(\frac{\pi R}{d \Gamma_0}
    e^{-\frac{\pi}{2}\frac{E_0}{\Gamma_0} - \gamma}),
  \\
  \label{eq:NearfieldAntisym}
  \xi_2(d) &=
  \frac{2}{\pi} \Gamma_0 W_0\qty(\frac{\pi R}{d \Gamma_0}
    e^{\frac{\pi}{2}\frac{E_0}{\Gamma_0} - \gamma}),
\end{align}
\end{subequations}
where \(W_n(\xi)\) is the $n$-th branch of the complex Lambert $W$~function~\cite{corless1996lambert}.  We note that \(\Im{\xi_1} = -2\Gamma_0\), \(\Im{\xi_2} = 0\) and \(\Re{\xi_1} < \Re{\xi_2}\).  The root \(\xi_1\) therefore corresponds to the symmetric eigenmode, and \(\xi_2\) corresponds to the antisymmetric one.

These solutions are plotted with dotted lines labeled `1st' in Fig.~\ref{fig:Eigenmodes}b. The real parts of the eigenfrequencies depend linearly on the logarithm of the distance $d$, while the imaginary parts remain constant. This behavior agrees well with the rigorous solution in the coupled oscillator model in the $d \to 0$ limit.
 
We find that the coupling coefficient is described well by its nearfield asymptotic before the first crossing of the real parts of the eigenfrequencies, and by the farfield asymptotic after the crossing, i.\,e. the crossing point is described by both of them. Thus, asymptotic approximations allow estimation of the critical distance for the transition between weak and strong coupling regimes.  Expressions~\eqref{eq:NearfieldSym} and~\eqref{eq:NearfieldAntisym} obtained in the nearfield asymptotic allow us to derive an estimation formula for the critical distance \(d_c\). To do that, we recall that for the antisymmetric mode $\Im{\xi_2(d)} \equiv 0$, and that at the crossing point, $\Re{\xi_2(d_c)} = E_0$. By substituting these into Eq.~\eqref{eq:NearfieldAntisym} we obtain $d_c$ as
\begin{equation}
  \label{eq:CrossingEstimate}
  d_c = \frac{2e^{-\gamma}}{E_0(\varepsilon)}R.
\end{equation}
Fig.~\ref{fig:Eigenmodes}b demonstrates that this formula slightly overestimates the crossing distance compared to the rigorous solution of the coupled oscillator model.

\begin{figure}[t]
\includegraphics[]{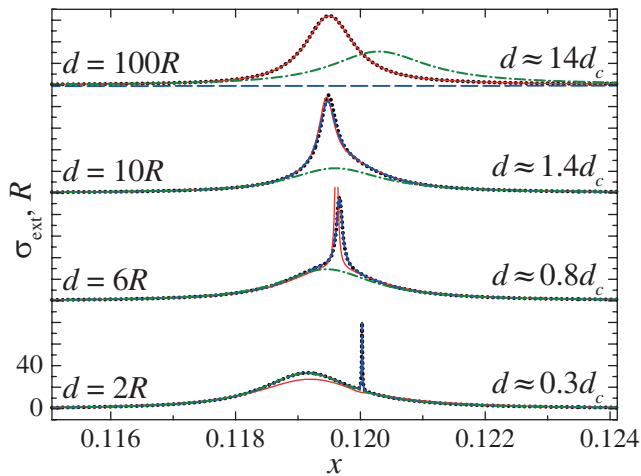}
\caption{\label{fig:SpectraApprox} 
Scattering spectra of two rods with $\varepsilon = 400$, calculated for TE~polarization within different approximations: dipole approximation (black dotted line), farfield asymptotic (red solid line), nearfield asymptotic truncated at the first term (green dash-dot line), and at the second term (blue dashed line). The critical distance $d_c$ here was calculated according to Eq.~\eqref{eq:ConvCoupling}.
}
\end{figure}

We have also simulated the scattering on the two rods with $\varepsilon = 400$ in the coupled oscillator model with the coupling approximated as the nearfield asymptotic. We plot the result with the dash-dot green line in Fig.~\ref{fig:SpectraApprox}. As is seen from the figure, for $d \le d_c$ the spectra obtained in this approximation do not exhibit the peak corresponding to the antisymmetric mode.  The reason is that the $Q$ factor of the antisymmetric mode frequency is infinitely large within this approximation. 

In order to correct that, we have solved numerically the equation for eigenfrequencies using the asymptotic expansion of the Hankel function up to the second term. We plot the result with dotted lines labeled `2nd' in Fig.~\ref{fig:Eigenmodes}b. These results agree with the rigorous solution up to $d = 4R$. For distances $d > 10R$ this approximation is no longer valid. We have also simulated the scattering on the two rods within this approximation. The result is shown with the blue dashed curve in Fig.~\ref{fig:SpectraApprox}. This curve shows good agreement with the exact solution (the black dotted line) for the distances up to $1.4d_c$. At larger distances both of the nearfield approximations fail to describe the spectra.

\subsubsection{Farfield asymptotic}

Now we consider the asymptotic expansion at $\xi \to \infty$ (see, e.g., Ref.~\onlinecite{watson1995treatise}):
\begin{equation}
  \label{eq:Farfield}
  H_0^{(1)}(\xi) \simeq \sqrt{\frac{2}{\pi \xi}} 
  e^{i\left(\xi - \frac{\pi}{4}\right)}.
\end{equation}
This corresponds to the farfield coupling. In this regime the coupling is carried by quasi-free waves described by Eq.~\eqref{eq:Farfield}. 

We substitute this asymptotic expansion into Eq.~\eqref{eq:CoupledOscillatorsEigenfreqs} and find the eigenfrequencies numerically. The results are plotted with dashed lines in Fig.~\ref{fig:Eigenmodes}b. As it is seen from the figure, the eigenmodes of the farfield-coupled system show good agreement with the rigorously computed eigenmodes for the distances greater than $d = 4R$.  The Fabry--Perot-like modes are also present in the farfield solution (they are not shown on the figure for clarity).  Therefore, the coupling of the rods is carried by quasi-free waves for distances $d > 4R$.

We also note that the Fabry--Perot-like eigenmodes should obey the phase synchronism relation
\begin{equation}
  \label{eq:FabryPerotPhase}
  2\left(d - \frac{\pi}{4}\right) + 2\arg\frac{\xi - E_0
  + i\Gamma_0}{-i\Gamma_0} = 2\pi N,
\end{equation}
which requires that the summary phase shift obtained by the quasi-free wave while it travels from the first rod to the second one, scatters on the second rod, then travels back to the first rod and scatters there once again, is equal to an integer $N$ multiplied by $2\pi$. It is straightforward that the equation~\eqref{eq:FabryPerotPhase} is a necessary condition for the zeros of the determinant of the coupled oscillator matrix within the farfield asymptotic.  This means that the eigenmodes that we have been referring to as ``Fabry--Perot-like eigenmodes'' can indeed be attributed to the Fabry--Perot resonances.

We have simulated the scattering on the two rods with $\varepsilon = 400$ with the coupling approximated as the farfield asymptotic. We plot the results with the red solid line in Fig.~\ref{fig:SpectraApprox}. As is seen from the figure, the spectra obtained in the farfield asymptotic agree well with the exact solution for $d \ge d_c$. At $d = 0.8d_c$ the farfield approximation overestimates the extinction cross-section while underestimating the Rabi splitting. At $d = 0.3d_c$ the spectrum obtained in the farfield approximation does not demonstrate the peak splitting effect at all.
 
\section{Conventional model}

In order to treat the Hamiltonian~\eqref{eq:CoupledOscillators} in the framework of the conventional theory of coupled oscillators~\cite{cao2015dielectric, limonov2017fano}, we neglect the frequency dependence of the coupling constant.  As is seen from Fig.~\ref{fig:Eigenmodes}a, dimer eigenfrequencies revolve around the Mie resonance frequency.  Moreover, Fig.~\ref{fig:Eigenmodes}b shows that the regions where the eigenmodes are described well by the farfield and the nearfield asymptotics, overlap at $d \simeq d_c$.  Hence, here we take the value of the coupling at the Mie resonance: $g = -i\Gamma_0H^{(1)}_0(d (E_0 - i\Gamma_0)/R)$. By doing so, we reduce the transcendental equation~\eqref{eq:CoupledOscillatorsEigenfreqs} to a quadratic one, which yields two eigenfrequencies
\begin{equation}
  \label{eq:GConstant}
  \xi_{1,2} = E_0 - i\Gamma_0 \pm g.
\end{equation}

In the conventional theory~\cite{cao2015dielectric, limonov2017fano}, where the coupling constant is real, the strong coupling regime corresponds to the case where the imaginary parts $\Gamma_{1,2}$ of $\xi_{1,2}$ coincide and the real parts $E_{1,2}$ are split. The reverse situation corresponds to the weak coupling regime.  This criterion may be formulated as follows:
\begin{equation}
  \label{eq:ConventionalCriterion}
  \begin{aligned}
    \abs{\Gamma_{1} - \Gamma_{2}} = 0 &\quad\mathrm{for~strong~coupling,} \\
    \abs{E_{1} - E_{2}} = 0 &\quad\mathrm{for~weak~coupling.}
  \end{aligned}
\end{equation}

However, here the oscillators are coupled via the continuum, so the coupling constant is complex and the Hamiltonian~\eqref{eq:CoupledOscillators} is non-Hermitian~\cite{cao2015dielectric, grimaudo2018exactly, joshi2018exceptional}.  Such systems are characterized by avoided crossings of the eigenfrequencies~\cite{cao2015dielectric}. The conventional criterion for the strong coupling does not work for these systems.  In order to distinguish between weak and strong coupling regimes in a non-Hermitian system, we generalize the conventional criterion~\eqref{eq:ConventionalCriterion} to the case of a complex coupling constant as follows:
\begin{subequations}
  \begin{align}
    \label{eq:OurStrongCoupling}
    \abs{\Gamma_{1} - \Gamma_{2}} \ll \abs{E_{1} - E_{2}} &\quad\mathrm{for~strong~coupling,} \\
    \abs{E_{1} - E_{2}} \ll \abs{\Gamma_{1} - \Gamma_{2}} &\quad\mathrm{for~weak~coupling.}
  \end{align}
\end{subequations}
We note that the inequality~\eqref{eq:OurStrongCoupling} holds true when $dE_0/R = j_{0,n}$, where $j_{0,n}$ are the zeros of the Bessel function $J_0(x)$. In other words, there exist infinitely large distances at which the inequality~\eqref{eq:OurStrongCoupling} holds. But, as was concluded from the analysis of scattering spectra, infinitely large distances correspond to infinitesimal coupling between the rods.
Thus, the inequality~\eqref{eq:OurStrongCoupling} is necessary but not sufficient for the strong coupling regime.

With the eigenfrequencies given by Eq.~\eqref{eq:GConstant}, for a high-$Q$ Mie resonance ($\Gamma_0 \ll E_0$) we can write
\begin{subequations}
  \begin{align}
    \abs{E_{1} - E_{2}} &\simeq \abs{\Gamma_0 Y_0 \qty(dE_0/R)}, \\
    \abs{\Gamma_{1} - \Gamma_{2}} &\simeq \abs{\Gamma_0 J_0\qty(dE_0/R)}.
  \end{align}
\end{subequations}
The inequality~\eqref{eq:OurStrongCoupling} for the strong coupling regime is therefore guaranteed to hold true with $dE_0/R \ll y_{0,1}$, where $y_{0,1} \simeq 0.89$ is the first zero of the Neumann function $Y_0(x)$. Thus, we can estimate the critical distance as
\begin{equation}
  d_c = \frac{y_{0,1}}{E_0(\varepsilon)}R.
  \label{eq:ConvCoupling}
\end{equation}
We plot the criterion (\ref{eq:ConvCoupling}) in Fig.~\ref{fig:CriteriaComparison} and find it in a good agreement with other criteria discussed above.

\section{Conclusion}

To summarize, we have studied the non-Hermitian problem of optical mode coupling in two dielectric rods. A number of approaches have been applied to this problem, which have been shown to give similar results. The critical distance for the weak-to-strong coupling transition (shown in Fig.~\ref{fig:CriteriaComparison}) has almost linear dependence on rod refractive index for both TE and TM polarizations. We have uncovered that the strong coupling regime is due to the non-radiative exchange of energy in the nearfield wave zone. Besides, we found a Fabry--Perot like coupling owing to the farfield scattering. 

The strong coupling regime due to the hybridization of resonances in parts of a complex system enables new effects being a subject of intensive studies during the last decade. In the present work we have comprehensively analyzed the simplest system and have proposed several criteria for evaluating critical distance. We have shown in Fig.~\ref{fig:CriteriaComparison} that the strong coupling regime could be achieved in pairs of dielectric rods made of optical materials, such as Si and Ge~\cite{baranov2017all} and \mbox{Ge$_2$Sb$_2$Te$_5$} \cite{wuttig2017phase}. Our predictions agree well with the results of the experiment performed on Si nanowires~\cite{cao2011optical}. We believe that the obtained dependences with minor corrections still work for the more complicated systems such several-particle oligomers and even periodic lattices in 2D and 3D.

\acknowledgements 

We thank M.\,F.~Limonov for fruitful discussions. This work was supported by the Ministry of Education and Science of the Russian Federation (Project 3.1500.2017/4.6).

\appendix
\section{Time-averaged energy flux}
\label{sec:AppendixA}

2D waves emitted by a dipole oriented along axis $z$ (we consider TE~polarization as an example) can be written as:
\begin{subequations}
\begin{align}
  \label{eq:EnergyE}
  \mathbf{E}(t) &= A_0 H^{(1)}_0\qty(k r) e^{-i\omega t} \mathbf{e}_z
  \\
  \label{eq:EnergyH}
  \mathbf{H}(t) &= \frac{-i}{\omega\mu\mu_0}
  A_0 k H^{(1)}_{1}\qty(k r) e^{-i\omega t}
  \mathbf{e}_{\phi}
  \\
  \label{eq:EnergyS}
  \mathbf{S}(t) &= -\Re{E_z}\Re{H_{\phi}}\mathbf{e}_r.
\end{align}
\end{subequations}
Thus, it generates only a radial component of the Poynting vector. Its time average is equal to
\begin{equation}
  \label{eq:Sr}
  \expval{S_r} = -\frac{1}{2}\Re{E_z H_{\phi}^{*}}.
\end{equation}
By substituting the dipole field~\eqref{eq:EnergyE} and~\eqref{eq:EnergyH}, we get
\begin{equation}
  \expval{S_r} 
  = \frac{\kappa}{\omega\mu\mu_0} \frac{\abs{A_0}^2}{2}
  \qty(
    Y_0\qty(k r)J_1\qty(k r)
    -
    Y_1\qty(k r)J_0\qty(k r)
  ).
\end{equation}
Bessel functions satisfy the following identity~\cite{watson1995treatise}:
\begin{equation}
  J_l(x)Y'_l(x) - J'_l(x)Y_l(x) \equiv \frac{2}{\pi x}.
\end{equation}
By evaluating it for \(l = 0\), we get
\begin{equation}
  \label{eq:RadiantFlux}
  \expval{S_r} =
  \abs{A_0}^2
  \frac{k }{\omega \mu\mu_0}
  \frac{1}{\pi k r}
\end{equation}
at any distance from the source.

On the other hand, in the farfield asymptotic the expressions for the fields of a dipole are as follows:
\begin{equation}
\begin{aligned}
  \Re (E_z) &=
  \abs{A_0} \sqrt{\frac{2}{\pi k r}}
  \cos\qty(k r - \omega t - \frac{\pi}{4} + \arg{A_0}),
  \\
  \Re (H_{\phi}) &=
  \frac{-k}{\omega\mu\mu_0}
  \abs{A_0} \sqrt{\frac{2}{\pi k r}} \cos\qty(k r - \omega t
  - \frac{\pi}{4} + \arg{A_0}),
\end{aligned}
\end{equation}
By substituting them into~\eqref{eq:Sr}, we get
\begin{equation}
  \label{eq:SrFarfield}
  S_{r}(t) =
  \abs{A_0}^2
  \frac{k }{\omega \mu\mu_0}
  \cos^2\qty(k r - \omega t - \frac{\pi}{4} + \arg{A_0})
  \frac{2}{\pi k r}. \\
\end{equation}
Averaging this result over time gives Eq.~\eqref{eq:RadiantFlux}.  Thus, the time-average of the energy flux from a dipole source equals that of its farfield component.  We also note from Eq.~\eqref{eq:SrFarfield} that $S_r$ is non-negative, so for the far field $\langle\abs{S_r(t)}\rangle = \abs{\langle{S_r(t)}\rangle}$. Thus, the tidal and time-averaged fluxes are equal.


\begin{thebibliography}{49}%
\makeatletter
\providecommand \@ifxundefined [1]{%
 \@ifx{#1\undefined}
}%
\providecommand \@ifnum [1]{%
 \ifnum #1\expandafter \@firstoftwo
 \else \expandafter \@secondoftwo
 \fi
}%
\providecommand \@ifx [1]{%
 \ifx #1\expandafter \@firstoftwo
 \else \expandafter \@secondoftwo
 \fi
}%
\providecommand \natexlab [1]{#1}%
\providecommand \enquote  [1]{``#1''}%
\providecommand \bibnamefont  [1]{#1}%
\providecommand \bibfnamefont [1]{#1}%
\providecommand \citenamefont [1]{#1}%
\providecommand \href@noop [0]{\@secondoftwo}%
\providecommand \href [0]{\begingroup \@sanitize@url \@href}%
\providecommand \@href[1]{\@@startlink{#1}\@@href}%
\providecommand \@@href[1]{\endgroup#1\@@endlink}%
\providecommand \@sanitize@url [0]{\catcode `\\12\catcode `\$12\catcode
  `\&12\catcode `\#12\catcode `\^12\catcode `\_12\catcode `\%12\relax}%
\providecommand \@@startlink[1]{}%
\providecommand \@@endlink[0]{}%
\providecommand \url  [0]{\begingroup\@sanitize@url \@url }%
\providecommand \@url [1]{\endgroup\@href {#1}{\urlprefix }}%
\providecommand \urlprefix  [0]{URL }%
\providecommand \Eprint [0]{\href }%
\providecommand \doibase [0]{http://dx.doi.org/}%
\providecommand \selectlanguage [0]{\@gobble}%
\providecommand \bibinfo  [0]{\@secondoftwo}%
\providecommand \bibfield  [0]{\@secondoftwo}%
\providecommand \translation [1]{[#1]}%
\providecommand \BibitemOpen [0]{}%
\providecommand \bibitemStop [0]{}%
\providecommand \bibitemNoStop [0]{.\EOS\space}%
\providecommand \EOS [0]{\spacefactor3000\relax}%
\providecommand \BibitemShut  [1]{\csname bibitem#1\endcsname}%
\let\auto@bib@innerbib\@empty
\bibitem [{\citenamefont {Harris}(2010)}]{harris2010emergence}%
  \BibitemOpen
  \bibfield  {author} {\bibinfo {author} {\bibfnamefont {S.}~\bibnamefont
  {Harris}},\ }\href@noop {} {\bibfield  {journal} {\bibinfo  {journal} {Nature
  Photon.}\ }\textbf {\bibinfo {volume} {4}},\ \bibinfo {pages} {748} (\bibinfo
  {year} {2010})}\BibitemShut {NoStop}%
\bibitem [{\citenamefont {Novotny}\ and\ \citenamefont {van
  Hulst}(2011)}]{novotny2011antennas}%
  \BibitemOpen
  \bibfield  {author} {\bibinfo {author} {\bibfnamefont {L.}~\bibnamefont
  {Novotny}}\ and\ \bibinfo {author} {\bibfnamefont {N.}~\bibnamefont {van
  Hulst}},\ }\href@noop {} {\bibfield  {journal} {\bibinfo  {journal} {Nature
  Photon.}\ }\textbf {\bibinfo {volume} {5}},\ \bibinfo {pages} {83} (\bibinfo
  {year} {2011})}\BibitemShut {NoStop}%
\bibitem [{\citenamefont {Kuznetsov}\ \emph {et~al.}(2016)\citenamefont
  {Kuznetsov}, \citenamefont {Miroshnichenko}, \citenamefont {Brongersma},
  \citenamefont {Kivshar},\ and\ \citenamefont
  {Luk'yanchuk}}]{kuznetsov2016optically}%
  \BibitemOpen
  \bibfield  {author} {\bibinfo {author} {\bibfnamefont {A.~I.}\ \bibnamefont
  {Kuznetsov}}, \bibinfo {author} {\bibfnamefont {A.~E.}\ \bibnamefont
  {Miroshnichenko}}, \bibinfo {author} {\bibfnamefont {M.~L.}\ \bibnamefont
  {Brongersma}}, \bibinfo {author} {\bibfnamefont {Y.~S.}\ \bibnamefont
  {Kivshar}}, \ and\ \bibinfo {author} {\bibfnamefont {B.}~\bibnamefont
  {Luk'yanchuk}},\ }\href@noop {} {\bibfield  {journal} {\bibinfo  {journal}
  {Science}\ }\textbf {\bibinfo {volume} {354}},\ \bibinfo {pages} {aag2472}
  (\bibinfo {year} {2016})}\BibitemShut {NoStop}%
\bibitem [{\citenamefont {Filonov}\ \emph {et~al.}(2014)\citenamefont
  {Filonov}, \citenamefont {Slobozhanyuk}, \citenamefont {Krasnok},
  \citenamefont {Belov}, \citenamefont {Nenasheva}, \citenamefont {Hopkins},
  \citenamefont {Miroshnichenko},\ and\ \citenamefont
  {Kivshar}}]{filonov2014field}%
  \BibitemOpen
  \bibfield  {author} {\bibinfo {author} {\bibfnamefont {D.~S.}\ \bibnamefont
  {Filonov}}, \bibinfo {author} {\bibfnamefont {A.~P.}\ \bibnamefont
  {Slobozhanyuk}}, \bibinfo {author} {\bibfnamefont {A.~E.}\ \bibnamefont
  {Krasnok}}, \bibinfo {author} {\bibfnamefont {P.~A.}\ \bibnamefont {Belov}},
  \bibinfo {author} {\bibfnamefont {E.~A.}\ \bibnamefont {Nenasheva}}, \bibinfo
  {author} {\bibfnamefont {B.}~\bibnamefont {Hopkins}}, \bibinfo {author}
  {\bibfnamefont {A.~E.}\ \bibnamefont {Miroshnichenko}}, \ and\ \bibinfo
  {author} {\bibfnamefont {Y.~S.}\ \bibnamefont {Kivshar}},\ }\href@noop {}
  {\bibfield  {journal} {\bibinfo  {journal} {Appl. Phys. Lett.}\ }\textbf
  {\bibinfo {volume} {104}},\ \bibinfo {pages} {021104} (\bibinfo {year}
  {2014})}\BibitemShut {NoStop}%
\bibitem [{\citenamefont {Miroshnichenko}\ and\ \citenamefont
  {Kivshar}(2012)}]{miroshnichenko2012fano}%
  \BibitemOpen
  \bibfield  {author} {\bibinfo {author} {\bibfnamefont {A.~E.}\ \bibnamefont
  {Miroshnichenko}}\ and\ \bibinfo {author} {\bibfnamefont {Y.~S.}\
  \bibnamefont {Kivshar}},\ }\href@noop {} {\bibfield  {journal} {\bibinfo
  {journal} {Nano Lett.}\ }\textbf {\bibinfo {volume} {12}},\ \bibinfo {pages}
  {6459} (\bibinfo {year} {2012})}\BibitemShut {NoStop}%
\bibitem [{\citenamefont {Chandel}\ \emph {et~al.}(2019)\citenamefont
  {Chandel}, \citenamefont {Singh}, \citenamefont {Agrawal}, \citenamefont
  {K.A.}, \citenamefont {Gupta}, \citenamefont {Venugopal},\ and\ \citenamefont
  {Ghosh}}]{chandel2019mueller}%
  \BibitemOpen
  \bibfield  {author} {\bibinfo {author} {\bibfnamefont {S.}~\bibnamefont
  {Chandel}}, \bibinfo {author} {\bibfnamefont {A.~K.}\ \bibnamefont {Singh}},
  \bibinfo {author} {\bibfnamefont {A.}~\bibnamefont {Agrawal}}, \bibinfo
  {author} {\bibfnamefont {A.}~\bibnamefont {K.A.}}, \bibinfo {author}
  {\bibfnamefont {A.}~\bibnamefont {Gupta}}, \bibinfo {author} {\bibfnamefont
  {A.}~\bibnamefont {Venugopal}}, \ and\ \bibinfo {author} {\bibfnamefont
  {N.}~\bibnamefont {Ghosh}},\ }\href@noop {} {\bibfield  {journal} {\bibinfo
  {journal} {Opt. Commun.}\ }\textbf {\bibinfo {volume} {432}},\ \bibinfo
  {pages} {84} (\bibinfo {year} {2019})}\BibitemShut {NoStop}%
\bibitem [{\citenamefont {Yan}\ \emph {et~al.}(2015{\natexlab{a}})\citenamefont
  {Yan}, \citenamefont {Liu}, \citenamefont {Lin}, \citenamefont {Wang},
  \citenamefont {Chen}, \citenamefont {Wang},\ and\ \citenamefont
  {Yang}}]{yan2015magnetically}%
  \BibitemOpen
  \bibfield  {author} {\bibinfo {author} {\bibfnamefont {J.~H.}\ \bibnamefont
  {Yan}}, \bibinfo {author} {\bibfnamefont {P.}~\bibnamefont {Liu}}, \bibinfo
  {author} {\bibfnamefont {Z.~Y.}\ \bibnamefont {Lin}}, \bibinfo {author}
  {\bibfnamefont {H.}~\bibnamefont {Wang}}, \bibinfo {author} {\bibfnamefont
  {H.~J.}\ \bibnamefont {Chen}}, \bibinfo {author} {\bibfnamefont {C.~X.}\
  \bibnamefont {Wang}}, \ and\ \bibinfo {author} {\bibfnamefont {G.~W.}\
  \bibnamefont {Yang}},\ }\href@noop {} {\bibfield  {journal} {\bibinfo
  {journal} {Nature Commun.}\ }\textbf {\bibinfo {volume} {6}} (\bibinfo {year}
  {2015}{\natexlab{a}})}\BibitemShut {NoStop}%
\bibitem [{\citenamefont {Yan}\ \emph {et~al.}(2015{\natexlab{b}})\citenamefont
  {Yan}, \citenamefont {Liu}, \citenamefont {Lin}, \citenamefont {Wang},
  \citenamefont {Chen}, \citenamefont {Wang},\ and\ \citenamefont
  {Yang}}]{yan2015directional}%
  \BibitemOpen
  \bibfield  {author} {\bibinfo {author} {\bibfnamefont {J.}~\bibnamefont
  {Yan}}, \bibinfo {author} {\bibfnamefont {P.}~\bibnamefont {Liu}}, \bibinfo
  {author} {\bibfnamefont {Z.}~\bibnamefont {Lin}}, \bibinfo {author}
  {\bibfnamefont {H.}~\bibnamefont {Wang}}, \bibinfo {author} {\bibfnamefont
  {H.}~\bibnamefont {Chen}}, \bibinfo {author} {\bibfnamefont {C.}~\bibnamefont
  {Wang}}, \ and\ \bibinfo {author} {\bibfnamefont {G.}~\bibnamefont {Yang}},\
  }\href@noop {} {\bibfield  {journal} {\bibinfo  {journal} {{ACS} Nano}\
  }\textbf {\bibinfo {volume} {9}},\ \bibinfo {pages} {2968} (\bibinfo {year}
  {2015}{\natexlab{b}})}\BibitemShut {NoStop}%
\bibitem [{\citenamefont {Wang}\ \emph {et~al.}(2007)\citenamefont {Wang},
  \citenamefont {Brandl}, \citenamefont {Nordlander},\ and\ \citenamefont
  {Halas}}]{wang2007plasmonic}%
  \BibitemOpen
  \bibfield  {author} {\bibinfo {author} {\bibfnamefont {H.}~\bibnamefont
  {Wang}}, \bibinfo {author} {\bibfnamefont {D.~W.}\ \bibnamefont {Brandl}},
  \bibinfo {author} {\bibfnamefont {P.}~\bibnamefont {Nordlander}}, \ and\
  \bibinfo {author} {\bibfnamefont {N.~J.}\ \bibnamefont {Halas}},\ }\href@noop
  {} {\bibfield  {journal} {\bibinfo  {journal} {Acc. Chem. Res.}\ }\textbf
  {\bibinfo {volume} {40}},\ \bibinfo {pages} {53} (\bibinfo {year}
  {2007})}\BibitemShut {NoStop}%
\bibitem [{\citenamefont {Greybush}\ \emph {et~al.}(2017)\citenamefont
  {Greybush}, \citenamefont {Liberal}, \citenamefont {Malassis}, \citenamefont
  {Kikkawa}, \citenamefont {Engheta}, \citenamefont {Murray},\ and\
  \citenamefont {Kagan}}]{greybush2017plasmon}%
  \BibitemOpen
  \bibfield  {author} {\bibinfo {author} {\bibfnamefont {N.~J.}\ \bibnamefont
  {Greybush}}, \bibinfo {author} {\bibfnamefont {I.}~\bibnamefont {Liberal}},
  \bibinfo {author} {\bibfnamefont {L.}~\bibnamefont {Malassis}}, \bibinfo
  {author} {\bibfnamefont {J.~M.}\ \bibnamefont {Kikkawa}}, \bibinfo {author}
  {\bibfnamefont {N.}~\bibnamefont {Engheta}}, \bibinfo {author} {\bibfnamefont
  {C.~B.}\ \bibnamefont {Murray}}, \ and\ \bibinfo {author} {\bibfnamefont
  {C.~R.}\ \bibnamefont {Kagan}},\ }\href@noop {} {\bibfield  {journal}
  {\bibinfo  {journal} {{ACS} Nano}\ }\textbf {\bibinfo {volume} {11}},\
  \bibinfo {pages} {2917} (\bibinfo {year} {2017})}\BibitemShut {NoStop}%
\bibitem [{\citenamefont {Alonso-Gonzalez}\ \emph {et~al.}(2011)\citenamefont
  {Alonso-Gonzalez}, \citenamefont {Schnell}, \citenamefont {Sarriugarte},
  \citenamefont {Sobhani}, \citenamefont {Wu}, \citenamefont {Arju},
  \citenamefont {Khanikaev}, \citenamefont {Golmar}, \citenamefont {Albella},
  \citenamefont {Arzubiaga}, \citenamefont {Casanova}, \citenamefont {Hueso},
  \citenamefont {Nordlander}, \citenamefont {Shvets},\ and\ \citenamefont
  {Hillenbrand}}]{alonso-gonzalez2011real}%
  \BibitemOpen
  \bibfield  {author} {\bibinfo {author} {\bibfnamefont {P.}~\bibnamefont
  {Alonso-Gonzalez}}, \bibinfo {author} {\bibfnamefont {M.}~\bibnamefont
  {Schnell}}, \bibinfo {author} {\bibfnamefont {P.}~\bibnamefont
  {Sarriugarte}}, \bibinfo {author} {\bibfnamefont {H.}~\bibnamefont
  {Sobhani}}, \bibinfo {author} {\bibfnamefont {C.}~\bibnamefont {Wu}},
  \bibinfo {author} {\bibfnamefont {N.}~\bibnamefont {Arju}}, \bibinfo {author}
  {\bibfnamefont {A.}~\bibnamefont {Khanikaev}}, \bibinfo {author}
  {\bibfnamefont {F.}~\bibnamefont {Golmar}}, \bibinfo {author} {\bibfnamefont
  {P.}~\bibnamefont {Albella}}, \bibinfo {author} {\bibfnamefont
  {L.}~\bibnamefont {Arzubiaga}}, \bibinfo {author} {\bibfnamefont
  {F.}~\bibnamefont {Casanova}}, \bibinfo {author} {\bibfnamefont {L.~E.}\
  \bibnamefont {Hueso}}, \bibinfo {author} {\bibfnamefont {P.}~\bibnamefont
  {Nordlander}}, \bibinfo {author} {\bibfnamefont {G.}~\bibnamefont {Shvets}},
  \ and\ \bibinfo {author} {\bibfnamefont {R.}~\bibnamefont {Hillenbrand}},\
  }\href@noop {} {\bibfield  {journal} {\bibinfo  {journal} {Nano Lett.}\
  }\textbf {\bibinfo {volume} {11}},\ \bibinfo {pages} {3922} (\bibinfo {year}
  {2011})}\BibitemShut {NoStop}%
\bibitem [{\citenamefont {Cao}\ \emph {et~al.}(2011)\citenamefont {Cao},
  \citenamefont {Fan},\ and\ \citenamefont {Brongersma}}]{cao2011optical}%
  \BibitemOpen
  \bibfield  {author} {\bibinfo {author} {\bibfnamefont {L.}~\bibnamefont
  {Cao}}, \bibinfo {author} {\bibfnamefont {P.}~\bibnamefont {Fan}}, \ and\
  \bibinfo {author} {\bibfnamefont {M.~L.}\ \bibnamefont {Brongersma}},\
  }\href@noop {} {\bibfield  {journal} {\bibinfo  {journal} {Nano Lett.}\
  }\textbf {\bibinfo {volume} {11}},\ \bibinfo {pages} {1463} (\bibinfo {year}
  {2011})}\BibitemShut {NoStop}%
\bibitem [{\citenamefont {Gao}\ \emph {et~al.}(2018)\citenamefont {Gao},
  \citenamefont {Zhou}, \citenamefont {Shi}, \citenamefont {Guo},\ and\
  \citenamefont {Tong}}]{gao2018dark}%
  \BibitemOpen
  \bibfield  {author} {\bibinfo {author} {\bibfnamefont {Y.}~\bibnamefont
  {Gao}}, \bibinfo {author} {\bibfnamefont {N.}~\bibnamefont {Zhou}}, \bibinfo
  {author} {\bibfnamefont {Z.}~\bibnamefont {Shi}}, \bibinfo {author}
  {\bibfnamefont {X.}~\bibnamefont {Guo}}, \ and\ \bibinfo {author}
  {\bibfnamefont {L.}~\bibnamefont {Tong}},\ }\href@noop {} {\bibfield
  {journal} {\bibinfo  {journal} {Photon. Res.}\ }\textbf {\bibinfo {volume}
  {6}},\ \bibinfo {pages} {887} (\bibinfo {year} {2018})}\BibitemShut {NoStop}%
\bibitem [{\citenamefont {Bakker}\ \emph {et~al.}(2015)\citenamefont {Bakker},
  \citenamefont {Permyakov}, \citenamefont {Yu}, \citenamefont {Markovich},
  \citenamefont {Paniagua-Dom{\'{\i}}nguez}, \citenamefont {Gonzaga},
  \citenamefont {Samusev}, \citenamefont {Kivshar}, \citenamefont
  {Luk'yanchuk},\ and\ \citenamefont {Kuznetsov}}]{bakker2015magnetic}%
  \BibitemOpen
  \bibfield  {author} {\bibinfo {author} {\bibfnamefont {R.~M.}\ \bibnamefont
  {Bakker}}, \bibinfo {author} {\bibfnamefont {D.}~\bibnamefont {Permyakov}},
  \bibinfo {author} {\bibfnamefont {Y.~F.}\ \bibnamefont {Yu}}, \bibinfo
  {author} {\bibfnamefont {D.}~\bibnamefont {Markovich}}, \bibinfo {author}
  {\bibfnamefont {R.}~\bibnamefont {Paniagua-Dom{\'{\i}}nguez}}, \bibinfo
  {author} {\bibfnamefont {L.}~\bibnamefont {Gonzaga}}, \bibinfo {author}
  {\bibfnamefont {A.}~\bibnamefont {Samusev}}, \bibinfo {author} {\bibfnamefont
  {Y.}~\bibnamefont {Kivshar}}, \bibinfo {author} {\bibfnamefont
  {B.}~\bibnamefont {Luk'yanchuk}}, \ and\ \bibinfo {author} {\bibfnamefont
  {A.~I.}\ \bibnamefont {Kuznetsov}},\ }\href@noop {} {\bibfield  {journal}
  {\bibinfo  {journal} {Nano Lett.}\ }\textbf {\bibinfo {volume} {15}},\
  \bibinfo {pages} {2137} (\bibinfo {year} {2015})}\BibitemShut {NoStop}%
\bibitem [{\citenamefont {Albella}\ \emph {et~al.}(2013)\citenamefont
  {Albella}, \citenamefont {Poyli}, \citenamefont {Schmidt}, \citenamefont
  {Maier}, \citenamefont {Moreno}, \citenamefont {S{\'{a}}enz},\ and\
  \citenamefont {Aizpurua}}]{albella2013low}%
  \BibitemOpen
  \bibfield  {author} {\bibinfo {author} {\bibfnamefont {P.}~\bibnamefont
  {Albella}}, \bibinfo {author} {\bibfnamefont {M.~A.}\ \bibnamefont {Poyli}},
  \bibinfo {author} {\bibfnamefont {M.~K.}\ \bibnamefont {Schmidt}}, \bibinfo
  {author} {\bibfnamefont {S.~A.}\ \bibnamefont {Maier}}, \bibinfo {author}
  {\bibfnamefont {F.}~\bibnamefont {Moreno}}, \bibinfo {author} {\bibfnamefont
  {J.~J.}\ \bibnamefont {S{\'{a}}enz}}, \ and\ \bibinfo {author} {\bibfnamefont
  {J.}~\bibnamefont {Aizpurua}},\ }\href@noop {} {\bibfield  {journal}
  {\bibinfo  {journal} {J. Phys. Chem. C}\ }\textbf {\bibinfo {volume} {117}},\
  \bibinfo {pages} {13573} (\bibinfo {year} {2013})}\BibitemShut {NoStop}%
\bibitem [{\citenamefont {Boudarham}\ \emph {et~al.}(2014)\citenamefont
  {Boudarham}, \citenamefont {Abdeddaim},\ and\ \citenamefont
  {Bonod}}]{boudarham2014enhancing}%
  \BibitemOpen
  \bibfield  {author} {\bibinfo {author} {\bibfnamefont {G.}~\bibnamefont
  {Boudarham}}, \bibinfo {author} {\bibfnamefont {R.}~\bibnamefont
  {Abdeddaim}}, \ and\ \bibinfo {author} {\bibfnamefont {N.}~\bibnamefont
  {Bonod}},\ }\href@noop {} {\bibfield  {journal} {\bibinfo  {journal} {Appl.
  Phys. Lett.}\ }\textbf {\bibinfo {volume} {104}},\ \bibinfo {pages} {021117}
  (\bibinfo {year} {2014})}\BibitemShut {NoStop}%
\bibitem [{\citenamefont {Bachelier}\ \emph {et~al.}(2008)\citenamefont
  {Bachelier}, \citenamefont {Russier-Antoine}, \citenamefont {Benichou},
  \citenamefont {Jonin}, \citenamefont {{Del Fatti}}, \citenamefont
  {Vall{\'{e}}e},\ and\ \citenamefont {Brevet}}]{bachelier2008fano}%
  \BibitemOpen
  \bibfield  {author} {\bibinfo {author} {\bibfnamefont {G.}~\bibnamefont
  {Bachelier}}, \bibinfo {author} {\bibfnamefont {I.}~\bibnamefont
  {Russier-Antoine}}, \bibinfo {author} {\bibfnamefont {E.}~\bibnamefont
  {Benichou}}, \bibinfo {author} {\bibfnamefont {C.}~\bibnamefont {Jonin}},
  \bibinfo {author} {\bibfnamefont {N.}~\bibnamefont {{Del Fatti}}}, \bibinfo
  {author} {\bibfnamefont {F.}~\bibnamefont {Vall{\'{e}}e}}, \ and\ \bibinfo
  {author} {\bibfnamefont {P.-F.}\ \bibnamefont {Brevet}},\ }\href@noop {}
  {\bibfield  {journal} {\bibinfo  {journal} {Phys. Rev. Lett.}\ }\textbf
  {\bibinfo {volume} {101}},\ \bibinfo {pages} {197401} (\bibinfo {year}
  {2008})}\BibitemShut {NoStop}%
\bibitem [{\citenamefont {Gunnarsson}\ \emph {et~al.}(2005)\citenamefont
  {Gunnarsson}, \citenamefont {Rindzevicius}, \citenamefont {Prikulis},
  \citenamefont {Kasemo}, \citenamefont {Käll}, \citenamefont {Zou},\ and\
  \citenamefont {Schatz}}]{gunnarsson2005confined}%
  \BibitemOpen
  \bibfield  {author} {\bibinfo {author} {\bibfnamefont {L.}~\bibnamefont
  {Gunnarsson}}, \bibinfo {author} {\bibfnamefont {T.}~\bibnamefont
  {Rindzevicius}}, \bibinfo {author} {\bibfnamefont {J.}~\bibnamefont
  {Prikulis}}, \bibinfo {author} {\bibfnamefont {B.}~\bibnamefont {Kasemo}},
  \bibinfo {author} {\bibfnamefont {M.}~\bibnamefont {Käll}}, \bibinfo
  {author} {\bibfnamefont {S.}~\bibnamefont {Zou}}, \ and\ \bibinfo {author}
  {\bibfnamefont {G.~C.}\ \bibnamefont {Schatz}},\ }\href@noop {} {\bibfield
  {journal} {\bibinfo  {journal} {J. Phys. Chem. B}\ }\textbf {\bibinfo
  {volume} {109}},\ \bibinfo {pages} {1079} (\bibinfo {year}
  {2005})}\BibitemShut {NoStop}%
\bibitem [{\citenamefont {Tsai}\ \emph {et~al.}(2012)\citenamefont {Tsai},
  \citenamefont {Lin}, \citenamefont {Wu}, \citenamefont {Lin}, \citenamefont
  {Lu},\ and\ \citenamefont {Lee}}]{tsai2012plasmonic}%
  \BibitemOpen
  \bibfield  {author} {\bibinfo {author} {\bibfnamefont {C.-Y.}\ \bibnamefont
  {Tsai}}, \bibinfo {author} {\bibfnamefont {J.-W.}\ \bibnamefont {Lin}},
  \bibinfo {author} {\bibfnamefont {C.-Y.}\ \bibnamefont {Wu}}, \bibinfo
  {author} {\bibfnamefont {P.-T.}\ \bibnamefont {Lin}}, \bibinfo {author}
  {\bibfnamefont {T.-W.}\ \bibnamefont {Lu}}, \ and\ \bibinfo {author}
  {\bibfnamefont {P.-T.}\ \bibnamefont {Lee}},\ }\href@noop {} {\bibfield
  {journal} {\bibinfo  {journal} {Nano Lett.}\ }\textbf {\bibinfo {volume}
  {12}},\ \bibinfo {pages} {1648} (\bibinfo {year} {2012})}\BibitemShut
  {NoStop}%
\bibitem [{\citenamefont {Schaffernak}\ \emph {et~al.}(2018)\citenamefont
  {Schaffernak}, \citenamefont {Krug}, \citenamefont {Belitsch}, \citenamefont
  {Ga{\v{s}}pari{\'{c}}}, \citenamefont {Ditlbacher}, \citenamefont
  {Hohenester}, \citenamefont {Krenn},\ and\ \citenamefont
  {Hohenau}}]{schaffernak2018plasmonic}%
  \BibitemOpen
  \bibfield  {author} {\bibinfo {author} {\bibfnamefont {G.}~\bibnamefont
  {Schaffernak}}, \bibinfo {author} {\bibfnamefont {M.~K.}\ \bibnamefont
  {Krug}}, \bibinfo {author} {\bibfnamefont {M.}~\bibnamefont {Belitsch}},
  \bibinfo {author} {\bibfnamefont {M.}~\bibnamefont {Ga{\v{s}}pari{\'{c}}}},
  \bibinfo {author} {\bibfnamefont {H.}~\bibnamefont {Ditlbacher}}, \bibinfo
  {author} {\bibfnamefont {U.}~\bibnamefont {Hohenester}}, \bibinfo {author}
  {\bibfnamefont {J.~R.}\ \bibnamefont {Krenn}}, \ and\ \bibinfo {author}
  {\bibfnamefont {A.}~\bibnamefont {Hohenau}},\ }\href@noop {} {\bibfield
  {journal} {\bibinfo  {journal} {{ACS} Photon.}\ } (\bibinfo {year}
  {2018})}\BibitemShut {NoStop}%
\bibitem [{\citenamefont {Khurgin}(2015)}]{khurgin2015how}%
  \BibitemOpen
  \bibfield  {author} {\bibinfo {author} {\bibfnamefont {J.~B.}\ \bibnamefont
  {Khurgin}},\ }\href@noop {} {\bibfield  {journal} {\bibinfo  {journal}
  {Nature Nanotech.}\ }\textbf {\bibinfo {volume} {10}},\ \bibinfo {pages} {2}
  (\bibinfo {year} {2015})}\BibitemShut {NoStop}%
\bibitem [{\citenamefont {Geddes}(2016)}]{geddes2016reviews}%
  \BibitemOpen
  \bibinfo {editor} {\bibfnamefont {C.~D.}\ \bibnamefont {Geddes}},\ ed.,\
  \href@noop {} {\emph {\bibinfo {title} {Reviews in Plasmonics 2015}}}\
  (\bibinfo  {publisher} {Springer International Publishing},\ \bibinfo {year}
  {2016})\BibitemShut {NoStop}%
\bibitem [{\citenamefont {Fernandes}\ \emph {et~al.}(2018)\citenamefont
  {Fernandes}, \citenamefont {Carvalho}, \citenamefont {{a}es}, \citenamefont
  {Neves},\ and\ \citenamefont {de~Assis}}]{fernandes2018center}%
  \BibitemOpen
  \bibfield  {author} {\bibinfo {author} {\bibfnamefont {T.~F.~D.}\
  \bibnamefont {Fernandes}}, \bibinfo {author} {\bibfnamefont {C.~M. K.~C.}\
  \bibnamefont {Carvalho}}, \bibinfo {author} {\bibfnamefont {P.~S. S.~G.}\
  \bibnamefont {{a}es}}, \bibinfo {author} {\bibfnamefont {B.~R.~A.}\
  \bibnamefont {Neves}}, \ and\ \bibinfo {author} {\bibfnamefont {P.-L.}\
  \bibnamefont {de~Assis}},\ }\href@noop {} {\bibfield  {journal} {\bibinfo
  {journal} {J. Lightwave Technol.}\ }\textbf {\bibinfo {volume} {36}},\
  \bibinfo {pages} {1608} (\bibinfo {year} {2018})}\BibitemShut {NoStop}%
\bibitem [{\citenamefont {Li}\ \emph {et~al.}(2009)\citenamefont {Li},
  \citenamefont {Pernice},\ and\ \citenamefont {Tang}}]{li2009tunable}%
  \BibitemOpen
  \bibfield  {author} {\bibinfo {author} {\bibfnamefont {M.}~\bibnamefont
  {Li}}, \bibinfo {author} {\bibfnamefont {W.~H.~P.}\ \bibnamefont {Pernice}},
  \ and\ \bibinfo {author} {\bibfnamefont {H.~X.}\ \bibnamefont {Tang}},\
  }\href@noop {} {\bibfield  {journal} {\bibinfo  {journal} {Nature Photon.}\
  }\textbf {\bibinfo {volume} {3}},\ \bibinfo {pages} {464} (\bibinfo {year}
  {2009})}\BibitemShut {NoStop}%
\bibitem [{\citenamefont {Rybin}\ \emph {et~al.}(2015)\citenamefont {Rybin},
  \citenamefont {Filonov}, \citenamefont {Samusev}, \citenamefont {Belov},
  \citenamefont {Kivshar},\ and\ \citenamefont {Limonov}}]{rybin2015phase}%
  \BibitemOpen
  \bibfield  {author} {\bibinfo {author} {\bibfnamefont {M.~V.}\ \bibnamefont
  {Rybin}}, \bibinfo {author} {\bibfnamefont {D.~S.}\ \bibnamefont {Filonov}},
  \bibinfo {author} {\bibfnamefont {K.~B.}\ \bibnamefont {Samusev}}, \bibinfo
  {author} {\bibfnamefont {P.~A.}\ \bibnamefont {Belov}}, \bibinfo {author}
  {\bibfnamefont {Y.~S.}\ \bibnamefont {Kivshar}}, \ and\ \bibinfo {author}
  {\bibfnamefont {M.~F.}\ \bibnamefont {Limonov}},\ }\href@noop {} {\bibfield
  {journal} {\bibinfo  {journal} {Nature Commun.}\ }\textbf {\bibinfo {volume}
  {6}} (\bibinfo {year} {2015})}\BibitemShut {NoStop}%
\bibitem [{\citenamefont {Maslova}\ \emph {et~al.}(2018)\citenamefont
  {Maslova}, \citenamefont {Limonov},\ and\ \citenamefont
  {Rybin}}]{maslova2018dielectric}%
  \BibitemOpen
  \bibfield  {author} {\bibinfo {author} {\bibfnamefont {E.~E.}\ \bibnamefont
  {Maslova}}, \bibinfo {author} {\bibfnamefont {M.~F.}\ \bibnamefont
  {Limonov}}, \ and\ \bibinfo {author} {\bibfnamefont {M.~V.}\ \bibnamefont
  {Rybin}},\ }\href@noop {} {\bibfield  {journal} {\bibinfo  {journal} {Opt.
  Lett.}\ }\textbf {\bibinfo {volume} {43}},\ \bibinfo {pages} {5516} (\bibinfo
  {year} {2018})}\BibitemShut {NoStop}%
\bibitem [{\citenamefont {Cao}\ and\ \citenamefont
  {Wiersig}(2015)}]{cao2015dielectric}%
  \BibitemOpen
  \bibfield  {author} {\bibinfo {author} {\bibfnamefont {H.}~\bibnamefont
  {Cao}}\ and\ \bibinfo {author} {\bibfnamefont {J.}~\bibnamefont {Wiersig}},\
  }\href@noop {} {\bibfield  {journal} {\bibinfo  {journal} {Rev. Mod. Phys.}\
  }\textbf {\bibinfo {volume} {87}},\ \bibinfo {pages} {61} (\bibinfo {year}
  {2015})}\BibitemShut {NoStop}%
\bibitem [{\citenamefont {Peng}\ \emph {et~al.}(2014)\citenamefont {Peng},
  \citenamefont {Özdemir}, \citenamefont {Chen}, \citenamefont {Nori},\ and\
  \citenamefont {Yang}}]{peng2014what}%
  \BibitemOpen
  \bibfield  {author} {\bibinfo {author} {\bibfnamefont {B.}~\bibnamefont
  {Peng}}, \bibinfo {author} {\bibfnamefont {{\c{S}}.~K.}\ \bibnamefont
  {Özdemir}}, \bibinfo {author} {\bibfnamefont {W.}~\bibnamefont {Chen}},
  \bibinfo {author} {\bibfnamefont {F.}~\bibnamefont {Nori}}, \ and\ \bibinfo
  {author} {\bibfnamefont {L.}~\bibnamefont {Yang}},\ }\href@noop {} {\bibfield
   {journal} {\bibinfo  {journal} {Nature Commun.}\ }\textbf {\bibinfo {volume}
  {5}} (\bibinfo {year} {2014})}\BibitemShut {NoStop}%
\bibitem [{\citenamefont {Dietz}\ \emph {et~al.}(2007)\citenamefont {Dietz},
  \citenamefont {Friedrich}, \citenamefont {Metz}, \citenamefont {Miski-Oglu},
  \citenamefont {Richter}, \citenamefont {Sch{\"a}fer},\ and\ \citenamefont
  {Stafford}}]{dietz2007rabi}%
  \BibitemOpen
  \bibfield  {author} {\bibinfo {author} {\bibfnamefont {B.}~\bibnamefont
  {Dietz}}, \bibinfo {author} {\bibfnamefont {T.}~\bibnamefont {Friedrich}},
  \bibinfo {author} {\bibfnamefont {J.}~\bibnamefont {Metz}}, \bibinfo {author}
  {\bibfnamefont {M.}~\bibnamefont {Miski-Oglu}}, \bibinfo {author}
  {\bibfnamefont {A.}~\bibnamefont {Richter}}, \bibinfo {author} {\bibfnamefont
  {F.}~\bibnamefont {Sch{\"a}fer}}, \ and\ \bibinfo {author} {\bibfnamefont
  {C.~A.}\ \bibnamefont {Stafford}},\ }\href@noop {} {\bibfield  {journal}
  {\bibinfo  {journal} {Phys. Rev. E}\ }\textbf {\bibinfo {volume} {75}},\
  \bibinfo {pages} {027201} (\bibinfo {year} {2007})}\BibitemShut {NoStop}%
\bibitem [{\citenamefont {Form{\'{a}}nek}\ \emph {et~al.}(2004)\citenamefont
  {Form{\'{a}}nek}, \citenamefont {Lombard},\ and\ \citenamefont
  {Mare{\v{s}}}}]{formanek2004wave}%
  \BibitemOpen
  \bibfield  {author} {\bibinfo {author} {\bibfnamefont {J.}~\bibnamefont
  {Form{\'{a}}nek}}, \bibinfo {author} {\bibfnamefont {R.}~\bibnamefont
  {Lombard}}, \ and\ \bibinfo {author} {\bibfnamefont {J.}~\bibnamefont
  {Mare{\v{s}}}},\ }\href@noop {} {\bibfield  {journal} {\bibinfo  {journal}
  {Czech. J. Phys.}\ }\textbf {\bibinfo {volume} {54}},\ \bibinfo {pages} {289}
  (\bibinfo {year} {2004})}\BibitemShut {NoStop}%
\bibitem [{\citenamefont {Lombard}\ and\ \citenamefont
  {Mare{\v{s}}}(2009)}]{lombard2009many}%
  \BibitemOpen
  \bibfield  {author} {\bibinfo {author} {\bibfnamefont {R.}~\bibnamefont
  {Lombard}}\ and\ \bibinfo {author} {\bibfnamefont {J.}~\bibnamefont
  {Mare{\v{s}}}},\ }\href@noop {} {\bibfield  {journal} {\bibinfo  {journal}
  {Phys. Lett. A}\ }\textbf {\bibinfo {volume} {373}},\ \bibinfo {pages} {426}
  (\bibinfo {year} {2009})}\BibitemShut {NoStop}%
\bibitem [{\citenamefont {Lloyd}\ and\ \citenamefont
  {Smith}(1972)}]{lloyd1972multiple}%
  \BibitemOpen
  \bibfield  {author} {\bibinfo {author} {\bibfnamefont {P.}~\bibnamefont
  {Lloyd}}\ and\ \bibinfo {author} {\bibfnamefont {P.}~\bibnamefont {Smith}},\
  }\href@noop {} {\bibfield  {journal} {\bibinfo  {journal} {Adv. Phys.}\
  }\textbf {\bibinfo {volume} {21}},\ \bibinfo {pages} {69} (\bibinfo {year}
  {1972})}\BibitemShut {NoStop}%
\bibitem [{\citenamefont {Leung}\ and\ \citenamefont
  {Qiu}(1993)}]{leung1993multiple}%
  \BibitemOpen
  \bibfield  {author} {\bibinfo {author} {\bibfnamefont {K.~M.}\ \bibnamefont
  {Leung}}\ and\ \bibinfo {author} {\bibfnamefont {Y.}~\bibnamefont {Qiu}},\
  }\href@noop {} {\bibfield  {journal} {\bibinfo  {journal} {Phys. Rev. B}\
  }\textbf {\bibinfo {volume} {48}},\ \bibinfo {pages} {7767} (\bibinfo {year}
  {1993})}\BibitemShut {NoStop}%
\bibitem [{\citenamefont {Felbacq}\ \emph {et~al.}(1994)\citenamefont
  {Felbacq}, \citenamefont {Tayeb},\ and\ \citenamefont
  {Maystre}}]{felbacq1994scattering}%
  \BibitemOpen
  \bibfield  {author} {\bibinfo {author} {\bibfnamefont {D.}~\bibnamefont
  {Felbacq}}, \bibinfo {author} {\bibfnamefont {G.}~\bibnamefont {Tayeb}}, \
  and\ \bibinfo {author} {\bibfnamefont {D.}~\bibnamefont {Maystre}},\
  }\href@noop {} {\bibfield  {journal} {\bibinfo  {journal} {J. Opt. Soc. Am.
  A}\ }\textbf {\bibinfo {volume} {11}},\ \bibinfo {pages} {2526} (\bibinfo
  {year} {1994})}\BibitemShut {NoStop}%
\bibitem [{\citenamefont {Nicorovici}\ \emph {et~al.}(1995)\citenamefont
  {Nicorovici}, \citenamefont {McPhedran},\ and\ \citenamefont
  {Botten}}]{nicorovici1995photonic}%
  \BibitemOpen
  \bibfield  {author} {\bibinfo {author} {\bibfnamefont {N.~A.}\ \bibnamefont
  {Nicorovici}}, \bibinfo {author} {\bibfnamefont {R.~C.}\ \bibnamefont
  {McPhedran}}, \ and\ \bibinfo {author} {\bibfnamefont {L.~C.}\ \bibnamefont
  {Botten}},\ }\href@noop {} {\bibfield  {journal} {\bibinfo  {journal} {Phys.
  Rev. E}\ }\textbf {\bibinfo {volume} {52}},\ \bibinfo {pages} {1135}
  (\bibinfo {year} {1995})}\BibitemShut {NoStop}%
\bibitem [{\citenamefont {Tayeb}\ and\ \citenamefont
  {Enoch}(2004)}]{tayeb2004combined}%
  \BibitemOpen
  \bibfield  {author} {\bibinfo {author} {\bibfnamefont {G.}~\bibnamefont
  {Tayeb}}\ and\ \bibinfo {author} {\bibfnamefont {S.}~\bibnamefont {Enoch}},\
  }\href@noop {} {\bibfield  {journal} {\bibinfo  {journal} {J. Opt. Soc. Am.
  A}\ }\textbf {\bibinfo {volume} {21}},\ \bibinfo {pages} {1417} (\bibinfo
  {year} {2004})}\BibitemShut {NoStop}%
\bibitem [{\citenamefont {Marko{\v{s}}}(2016)}]{markos2016photonic}%
  \BibitemOpen
  \bibfield  {author} {\bibinfo {author} {\bibfnamefont {P.}~\bibnamefont
  {Marko{\v{s}}}},\ }\href@noop {} {\bibfield  {journal} {\bibinfo  {journal}
  {Opt. Commun.}\ }\textbf {\bibinfo {volume} {361}},\ \bibinfo {pages} {65}
  (\bibinfo {year} {2016})}\BibitemShut {NoStop}%
\bibitem [{\citenamefont {Marko{\v{s}}}\ and\ \citenamefont
  {Kuzmiak}(2016)}]{markos2016coupling}%
  \BibitemOpen
  \bibfield  {author} {\bibinfo {author} {\bibfnamefont {P.}~\bibnamefont
  {Marko{\v{s}}}}\ and\ \bibinfo {author} {\bibfnamefont {V.}~\bibnamefont
  {Kuzmiak}},\ }\href@noop {} {\bibfield  {journal} {\bibinfo  {journal} {Phys.
  Rev. A}\ }\textbf {\bibinfo {volume} {94}},\ \bibinfo {pages} {033845}
  (\bibinfo {year} {2016})}\BibitemShut {NoStop}%
\bibitem [{\citenamefont {Dmitriev}\ and\ \citenamefont
  {Rybin}(2018)}]{dmitriev2018coupling}%
  \BibitemOpen
  \bibfield  {author} {\bibinfo {author} {\bibfnamefont {A.~A.}\ \bibnamefont
  {Dmitriev}}\ and\ \bibinfo {author} {\bibfnamefont {M.~V.}\ \bibnamefont
  {Rybin}},\ }in\ \href@noop {} {\emph {\bibinfo {booktitle} {2018 Days on
  Diffraction (DD)}}}\ (\bibinfo {year} {2018})\ pp.\ \bibinfo {pages}
  {71--75}\BibitemShut {NoStop}%
\bibitem [{\citenamefont {Mitri}(2015)}]{mitri2015optical}%
  \BibitemOpen
  \bibfield  {author} {\bibinfo {author} {\bibfnamefont {F.}~\bibnamefont
  {Mitri}},\ }\href@noop {} {\bibfield  {journal} {\bibinfo  {journal}
  {Ultrasonics}\ }\textbf {\bibinfo {volume} {62}},\ \bibinfo {pages} {20}
  (\bibinfo {year} {2015})}\BibitemShut {NoStop}%
\bibitem [{\citenamefont {Gu}\ and\ \citenamefont {Qian}(1989)}]{gu1989some}%
  \BibitemOpen
  \bibfield  {author} {\bibinfo {author} {\bibfnamefont {Z.-Y.}\ \bibnamefont
  {Gu}}\ and\ \bibinfo {author} {\bibfnamefont {S.-W.}\ \bibnamefont {Qian}},\
  }\href@noop {} {\bibfield  {journal} {\bibinfo  {journal} {Phys. Lett. A}\
  }\textbf {\bibinfo {volume} {136}},\ \bibinfo {pages} {6} (\bibinfo {year}
  {1989})}\BibitemShut {NoStop}%
\bibitem [{\citenamefont {Andryieuski}\ \emph {et~al.}(2015)\citenamefont
  {Andryieuski}, \citenamefont {Kuznetsova}, \citenamefont {Zhukovsky},
  \citenamefont {Kivshar},\ and\ \citenamefont
  {Lavrinenko}}]{andryieuski2015water}%
  \BibitemOpen
  \bibfield  {author} {\bibinfo {author} {\bibfnamefont {A.}~\bibnamefont
  {Andryieuski}}, \bibinfo {author} {\bibfnamefont {S.~M.}\ \bibnamefont
  {Kuznetsova}}, \bibinfo {author} {\bibfnamefont {S.~V.}\ \bibnamefont
  {Zhukovsky}}, \bibinfo {author} {\bibfnamefont {Y.~S.}\ \bibnamefont
  {Kivshar}}, \ and\ \bibinfo {author} {\bibfnamefont {A.~V.}\ \bibnamefont
  {Lavrinenko}},\ }\href@noop {} {\bibfield  {journal} {\bibinfo  {journal}
  {Sci. Rep.}\ }\textbf {\bibinfo {volume} {5}} (\bibinfo {year}
  {2015})}\BibitemShut {NoStop}%
\bibitem [{\citenamefont {Baranov}\ \emph {et~al.}(2017)\citenamefont
  {Baranov}, \citenamefont {Zuev}, \citenamefont {Lepeshov}, \citenamefont
  {Kotov}, \citenamefont {Krasnok}, \citenamefont {Evlyukhin},\ and\
  \citenamefont {Chichkov}}]{baranov2017all}%
  \BibitemOpen
  \bibfield  {author} {\bibinfo {author} {\bibfnamefont {D.~G.}\ \bibnamefont
  {Baranov}}, \bibinfo {author} {\bibfnamefont {D.~A.}\ \bibnamefont {Zuev}},
  \bibinfo {author} {\bibfnamefont {S.~I.}\ \bibnamefont {Lepeshov}}, \bibinfo
  {author} {\bibfnamefont {O.~V.}\ \bibnamefont {Kotov}}, \bibinfo {author}
  {\bibfnamefont {A.~E.}\ \bibnamefont {Krasnok}}, \bibinfo {author}
  {\bibfnamefont {A.~B.}\ \bibnamefont {Evlyukhin}}, \ and\ \bibinfo {author}
  {\bibfnamefont {B.~N.}\ \bibnamefont {Chichkov}},\ }\href@noop {} {\bibfield
  {journal} {\bibinfo  {journal} {Optica}\ }\textbf {\bibinfo {volume} {4}},\
  \bibinfo {pages} {814} (\bibinfo {year} {2017})}\BibitemShut {NoStop}%
\bibitem [{\citenamefont {Wuttig}\ \emph {et~al.}(2017)\citenamefont {Wuttig},
  \citenamefont {Bhaskaran},\ and\ \citenamefont {Taubner}}]{wuttig2017phase}%
  \BibitemOpen
  \bibfield  {author} {\bibinfo {author} {\bibfnamefont {M.}~\bibnamefont
  {Wuttig}}, \bibinfo {author} {\bibfnamefont {H.}~\bibnamefont {Bhaskaran}}, \
  and\ \bibinfo {author} {\bibfnamefont {T.}~\bibnamefont {Taubner}},\
  }\href@noop {} {\bibfield  {journal} {\bibinfo  {journal} {Nat. Photon.}\
  }\textbf {\bibinfo {volume} {11}},\ \bibinfo {pages} {465} (\bibinfo {year}
  {2017})}\BibitemShut {NoStop}%
\bibitem [{\citenamefont {Watson}(1995)}]{watson1995treatise}%
  \BibitemOpen
  \bibfield  {author} {\bibinfo {author} {\bibfnamefont {G.~N.}\ \bibnamefont
  {Watson}},\ }\href@noop {} {\emph {\bibinfo {title} {A treatise on the theory
  of Bessel functions}}}\ (\bibinfo  {publisher} {Cambridge university press},\
  \bibinfo {year} {1995})\BibitemShut {NoStop}%
\bibitem [{\citenamefont {Corless}\ \emph {et~al.}(1996)\citenamefont
  {Corless}, \citenamefont {Gonnet}, \citenamefont {Hare}, \citenamefont
  {Jeffrey},\ and\ \citenamefont {Knuth}}]{corless1996lambert}%
  \BibitemOpen
  \bibfield  {author} {\bibinfo {author} {\bibfnamefont {R.~M.}\ \bibnamefont
  {Corless}}, \bibinfo {author} {\bibfnamefont {G.~H.}\ \bibnamefont {Gonnet}},
  \bibinfo {author} {\bibfnamefont {D.~E.~G.}\ \bibnamefont {Hare}}, \bibinfo
  {author} {\bibfnamefont {D.~J.}\ \bibnamefont {Jeffrey}}, \ and\ \bibinfo
  {author} {\bibfnamefont {D.~E.}\ \bibnamefont {Knuth}},\ }\href@noop {}
  {\bibfield  {journal} {\bibinfo  {journal} {Adv. in Comput. Math.}\ }\textbf
  {\bibinfo {volume} {5}},\ \bibinfo {pages} {329} (\bibinfo {year}
  {1996})}\BibitemShut {NoStop}%
\bibitem [{\citenamefont {Limonov}\ \emph {et~al.}(2017)\citenamefont
  {Limonov}, \citenamefont {Rybin}, \citenamefont {Poddubny},\ and\
  \citenamefont {Kivshar}}]{limonov2017fano}%
  \BibitemOpen
  \bibfield  {author} {\bibinfo {author} {\bibfnamefont {M.~F.}\ \bibnamefont
  {Limonov}}, \bibinfo {author} {\bibfnamefont {M.~V.}\ \bibnamefont {Rybin}},
  \bibinfo {author} {\bibfnamefont {A.~N.}\ \bibnamefont {Poddubny}}, \ and\
  \bibinfo {author} {\bibfnamefont {Y.~S.}\ \bibnamefont {Kivshar}},\
  }\href@noop {} {\bibfield  {journal} {\bibinfo  {journal} {Nature Photon.}\
  }\textbf {\bibinfo {volume} {11}},\ \bibinfo {pages} {543} (\bibinfo {year}
  {2017})}\BibitemShut {NoStop}%
\bibitem [{\citenamefont {Grimaudo}\ \emph {et~al.}(2018)\citenamefont
  {Grimaudo}, \citenamefont {de~Castro}, \citenamefont {Ku{\'{s}}},\ and\
  \citenamefont {Messina}}]{grimaudo2018exactly}%
  \BibitemOpen
  \bibfield  {author} {\bibinfo {author} {\bibfnamefont {R.}~\bibnamefont
  {Grimaudo}}, \bibinfo {author} {\bibfnamefont {A.~S.~M.}\ \bibnamefont
  {de~Castro}}, \bibinfo {author} {\bibfnamefont {M.}~\bibnamefont
  {Ku{\'{s}}}}, \ and\ \bibinfo {author} {\bibfnamefont {A.}~\bibnamefont
  {Messina}},\ }\href@noop {} {\bibfield  {journal} {\bibinfo  {journal} {Phys.
  Rev. A}\ }\textbf {\bibinfo {volume} {98}},\ \bibinfo {pages} {033835}
  (\bibinfo {year} {2018})}\BibitemShut {NoStop}%
\bibitem [{\citenamefont {Joshi}\ and\ \citenamefont
  {Galbraith}(2018)}]{joshi2018exceptional}%
  \BibitemOpen
  \bibfield  {author} {\bibinfo {author} {\bibfnamefont {S.}~\bibnamefont
  {Joshi}}\ and\ \bibinfo {author} {\bibfnamefont {I.}~\bibnamefont
  {Galbraith}},\ }\href@noop {} {\bibfield  {journal} {\bibinfo  {journal}
  {Phys. Rev. A}\ }\textbf {\bibinfo {volume} {98}},\ \bibinfo {pages} {042117}
  (\bibinfo {year} {2018})}\BibitemShut {NoStop}%
\end{thebibliography}
\end{document}